\documentclass[twocolumn,secnumarabic,amsmath, amssymb, nobibnotes, aps, pra, prd]{revtex4-2}
\usepackage{graphicx}

\begin{document}

\title{2D optical rogue waves affected by transverse carrier diffusion in broad-area semiconductor lasers with a saturable absorber}%

\author{K.~Talouneh$^1$}
\author{R.~Kheradmand$^{1,2}$}%
\author{G.~Tissoni$^3$}
\author{M.~Eslami$^4$}
\email[Corresponding author: ]{meslami@guilan.ac.ir}
\affiliation{$^1$ Faculty of Physics, University of Tabriz, Tabriz, Iran}
\affiliation{$^2$ Research Institute for Applied Physics and Astronomy, University of Tabriz, Tabriz, Iran}
\affiliation{$^3$ Universit\'{e} C\^{o}te \'{d}Azur, CNRS, Institut de Physique de Nice, Valbonne, France}
\affiliation{$^4$ Department of Physics, University of Guilan, P.O. Box 41335-1914 Rasht, Iran}


\begin{abstract}
Statistics and dynamics of 2D rogue waves in a broad-area semiconductor laser with an intracavity saturable absorber are numerically investigated under the effect of transverse carrier diffusion. We show that lateral diffusion of carriers alters the statistics of rogue waves by enhancing their formation in smaller ratios of carrier lifetimes in the active and passive materials while suppressing them when the ratio is larger. Temporal dynamics of the emitted rogue waves is also studied and shown that finite nonzero transverse carrier diffusion coefficient gives them a longer duration. To further approach the realistic experimental situation, we also investigated statistics and dynamics of rogue waves by simulating a circular disk-shape pump which replaces the flat pump profile typically used in numerical simulations of broad-area lasers. We show that finite pump shape reduces the number emitted rogue waves per unit area for large carrier lifetime ratios and increases that for smaller values of the ratio in both below and above laser threshold. Temporal width of the emitted rogue waves is also shown to reduce as a consequence of removing the nonphysical effects of infinite flat pump on carrier dynamics.
\end{abstract}

\maketitle
\section{Introduction}
\begin{figure}
	\includegraphics[width=1\columnwidth]{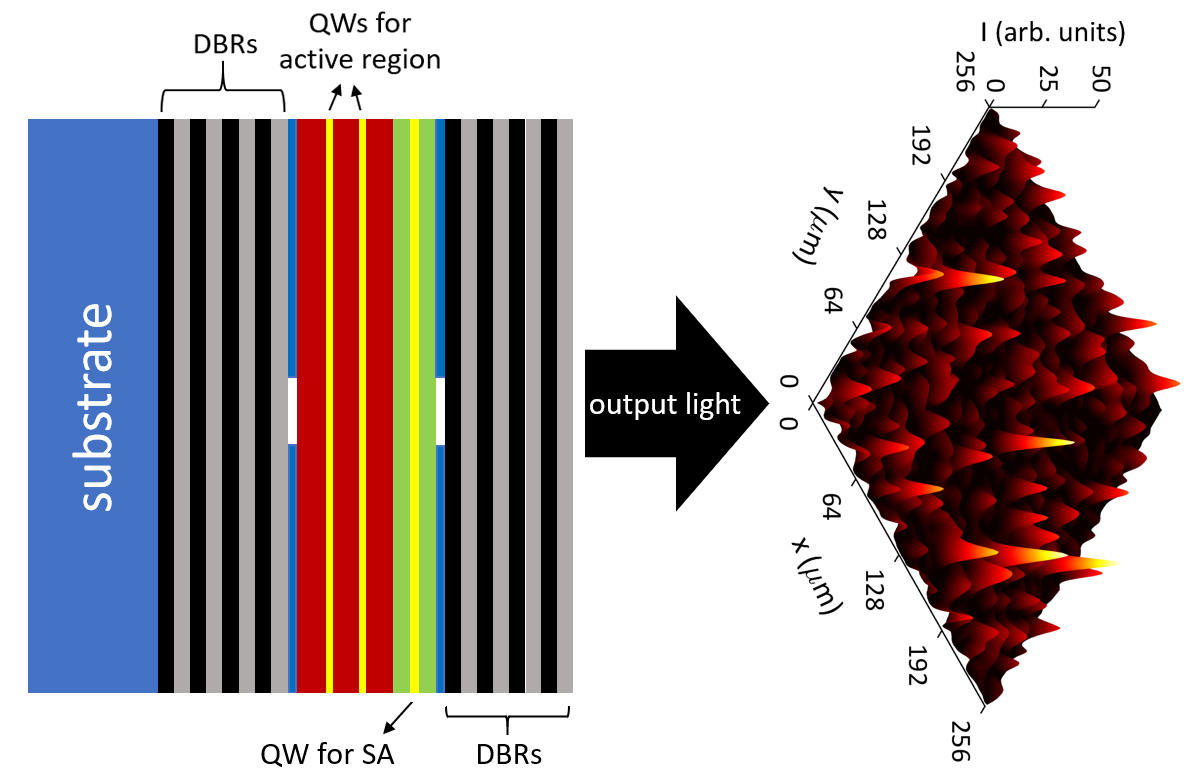}
	\caption{Schematic representation of a VCSEL with an intracavity saturable absorber. DBR, QW, and SA respectively stand for Distributed Bragg Reflector, Quantum Well, and Saturable Absorber.}
	\label{scheme}
\end{figure}

Semiconductor lasers, which constitute a major part of semiconductor photonics, are efficient devices and enabling technology that have opened many novel prospects in the modern information society from miniature semiconductor lasers which drive tech gadgets to the lasers that are employed in the modern communication systems. Their complex dynamics and nonlinear features, particularly in spatially extended systems such as Vertical-Cavity Surface Emitting Lasers (VCSELs), have attracted strong scientific interest as well as emerging technological significance \cite{VCSEL}. Most important of them have been the Cavity Solitons (CSs) and Light Bullets (LBs) studied in variety of configurations from injected semiconductor lasers \cite{Barbay11,Nature2002,IEEE06,McIntyre10,Prati10,Eslami462014,Eslami612014,Taghavi212018,Anbardan212019,Anbardan4742020} to the ones with delayed feedback \cite{Tlidi09,Panajotov10,Tlidi12,Garbin17,Pimenov18,Scroggie09} to those with intracavity saturable absorber \cite{CSL05,CSL07,Columbo10,Elsass10,Turconi15,Eslami892014,Eslami692015,Eslami192016,Eslami982018}. The semiconductor laser with an intracavity saturable absorber is the most interesting for applications due to its compactness and the possibility of integration in an all-optical circuit. However, fabrication of these devices poses the challenge of positioning the gain (pumped) and absorption (not pumped) elements as close as possible to allow for maximum interaction through the intracavity field. In a common design, the saturable absorber consisting of a quantum-well is placed either in the upper mirror stack or inside a second cavity coupled to the one containing the gain medium, see for example \cite{VCSELsa1,VCSELsa2}. An alternative approach has also been introduced where optical pumping is used which allows the coherent properties of the pump source to be used in the design of a compact and monolithic cavity for self-pulsing or bistable lasers \cite{Elsass10}. A schematic figure of the VCSEL with a saturable absorber used in our study is shown in Fig.~\ref{scheme}.\\
\begin{figure*}
	\includegraphics[width=0.6\columnwidth]{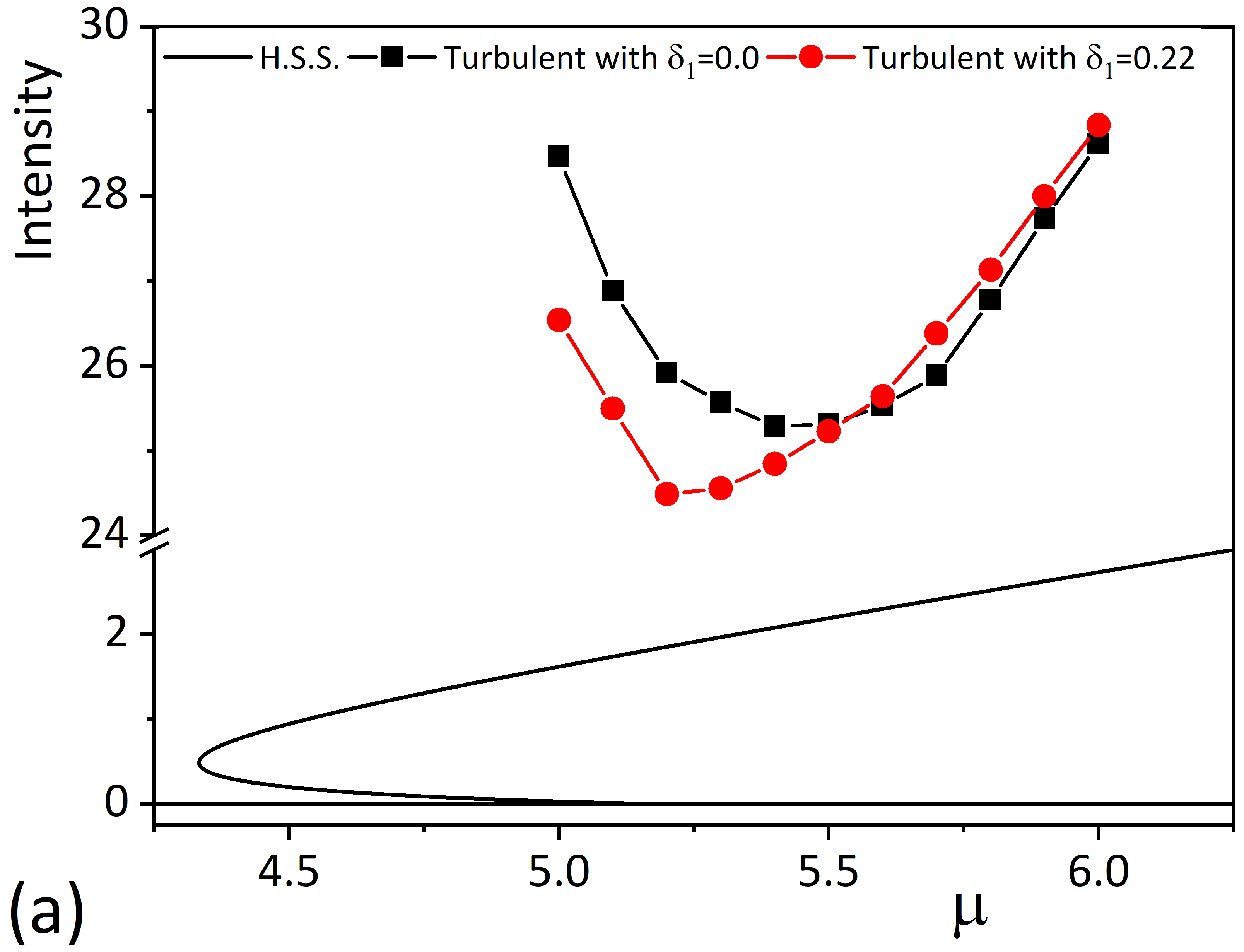}\quad \includegraphics[width=0.6\columnwidth]{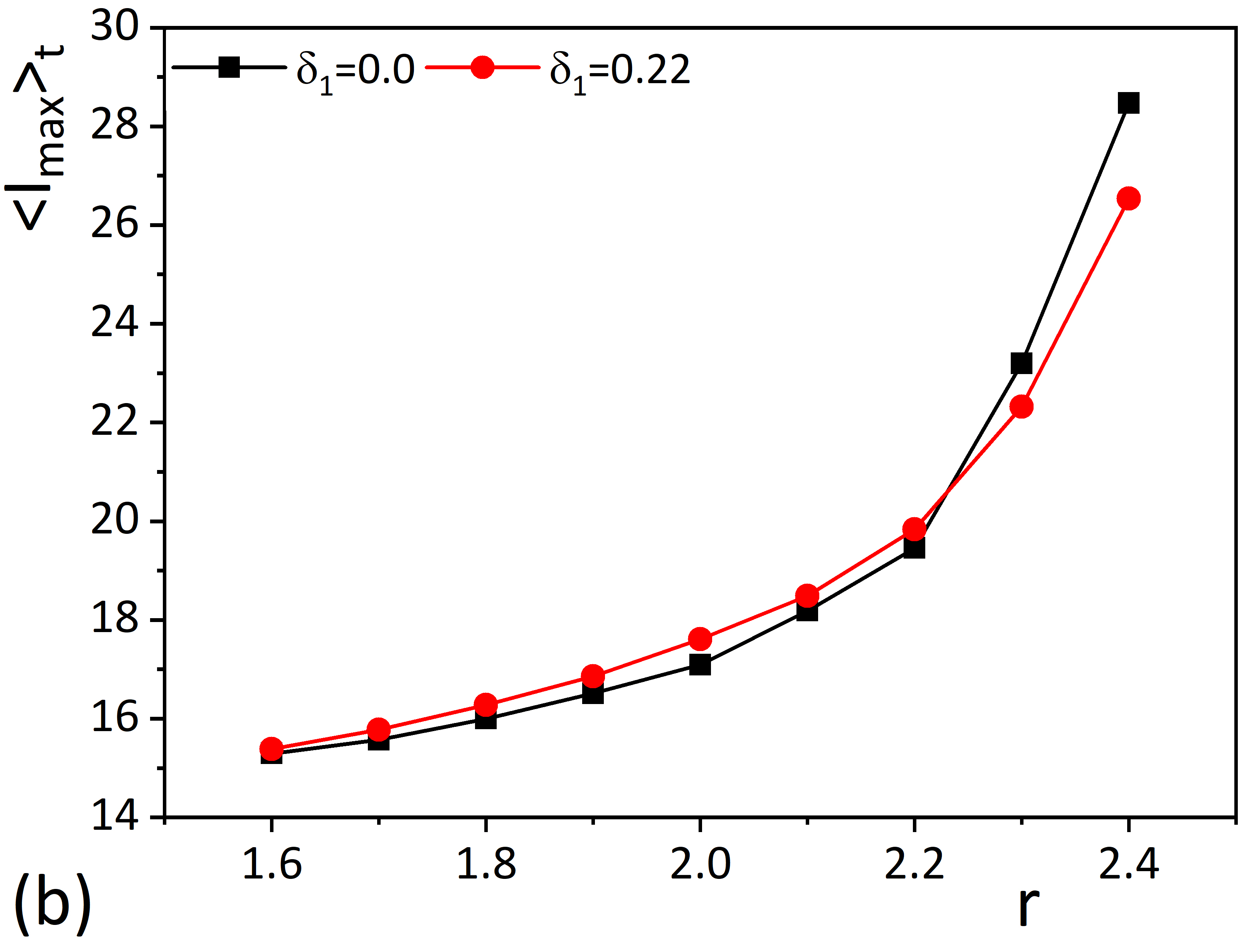} \quad \includegraphics[width=0.6\columnwidth]{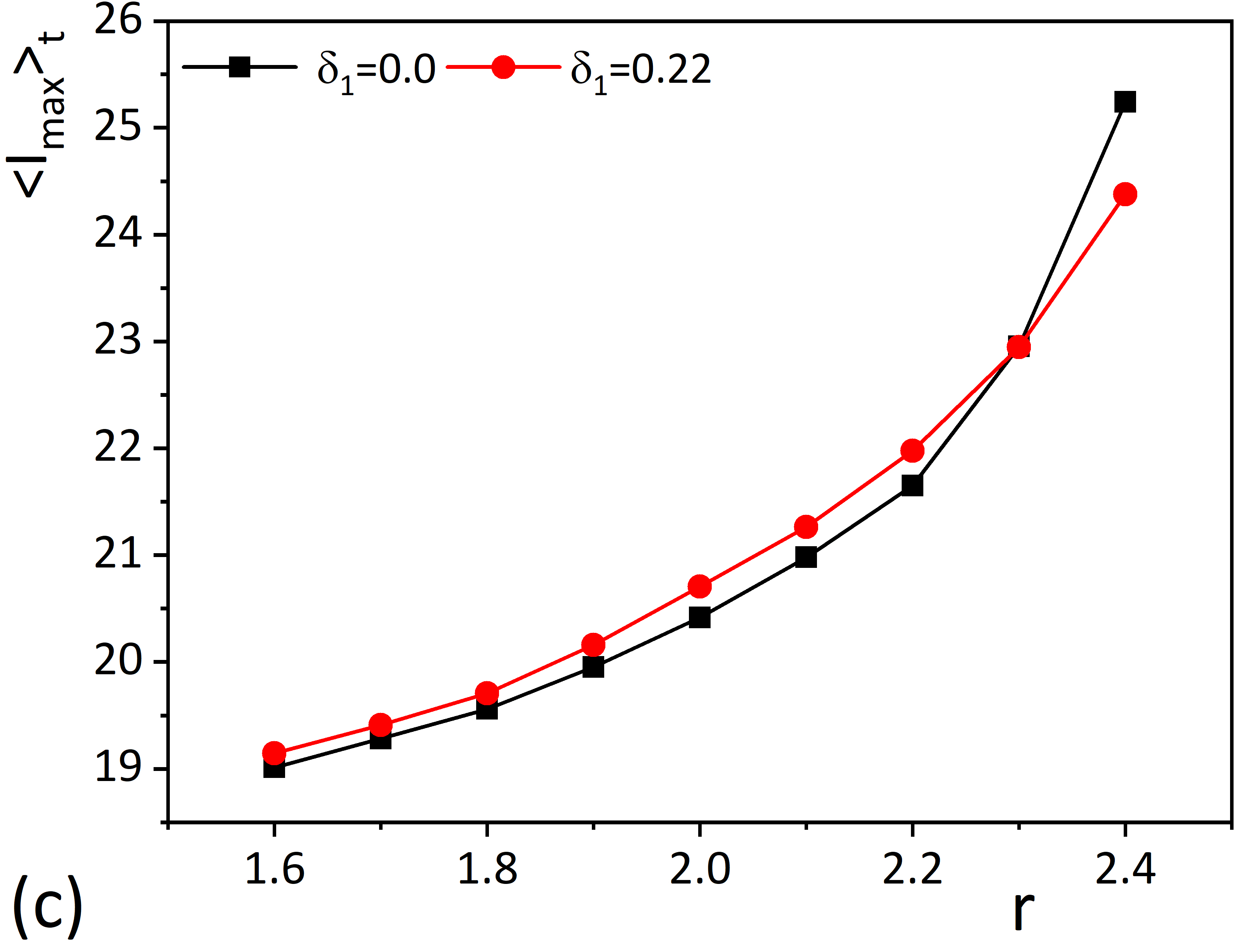}
	\caption{Turbulent branches for $\delta_1=0$ and $\delta_1=0.22$ in pump current $\mu$ and  $r$ scans. $r$ is fixed at 2.4 for (a); $\mu=5$ (below laser threshold) for (b) and $\mu=5.35$ (above laser threshold) for (c). For the turbulent branches in (a) we have depicted the time average of maximum intensities.}
	\label{branches}
\end{figure*}
Since more than a decade, investigation of rare and extreme pulses in optical systems has been a hot topic among researchers of nonlinear photonics \cite{Solli07,Akhmediev10,roadmap}. These extreme events in optical wave domain, referred to as rogue waves (RWs), are generally characterized by their probability of occurrence deviating from standard Gaussian statistical models and the peak height exceeding a certain system-dependent threshold which all roughly imply a value several times higher than the long-time average. Later on, spatiotemporal complexity associated with broad-area optical systems (both active and passive) inspired researchers to investigate 2D RWs \cite{Oppo13,Gibson16,Eslami17,Rimoldi17,Eslami20}. Statistics and dynamics of transverse RWs in broad-area semiconductor lasers with a saturable absorber were numerically studied recently in \cite{Rimoldi17,Eslami20}, for the underlying mechanisms and controlling their occurrences, and experimentally in \cite{Selmi16}.\\
Since broad-area semiconductor lasers have a broad stripe width of the active region ($\sim 200 \mu m$), the effects related to transverse carrier diffusion are important in the study of such a structure. The effect is minimum for the case where localized structures are considered but it is expected to play a significant role when extended turbulent structures are investigated. Here we focus on the effect of carrier transverse diffusion on the properties of the turbulent solutions and on the statistical and dynamical behavior of RWs in a broad-area semiconductor laser with an intracavity saturable absorber by slightly changing the model used in \cite{CSL07,Rimoldi17,Eslami20}. We extend the results in \cite{Rimoldi17,Eslami20} and show that inclusion of lateral carrier diffusion has a significant effect on the number of emitted RWs and their temporal dynamics. We particularly illustrate that transverse drift of carriers in a turbulent state eases its chaoticity to some extent and causes a reduction in the peak value of intensities and suppression of the RW formation for high values of carrier lifetimes ratio. However, RW emission is enhanced in nonzero carrier diffusion coefficients when the ratio of the carrier lifetimes in the two materials is decreased. Broader temporal width for RWs is also evidenced by the presence of nonzero transverse carrier diffusion coefficient.\\
To further approach the realistic view, we also studied the effects of a disk-shaped pump profile on RWs since in experimental conditions the pump shape controls the current-density distribution which is nonuniform across the transverse section. We show that when the carrier lifetimes ratio is lower, the number of emitted RWs increases in the presence of the finite circular pump. This is reversed for increased values of the carrier lifetimes ratio where the emission of RWs is suppressed by the effect of finite circular pump. Temporal dynamics of RWs is also discussed and shown that finite pump reduces their temporal width.\\
In section 2 model equations are detailed and the inclusion of transverse carrier diffusion coefficient is discussed. Section 3 and 4 are respectively devoted to the turbulent solutions and RWs under the effect of transverse carrier diffusion. Section 5 discusses the finite pump effects and conclusions are drawn in section 6.
\begin{figure}
	\includegraphics[width=1\columnwidth]{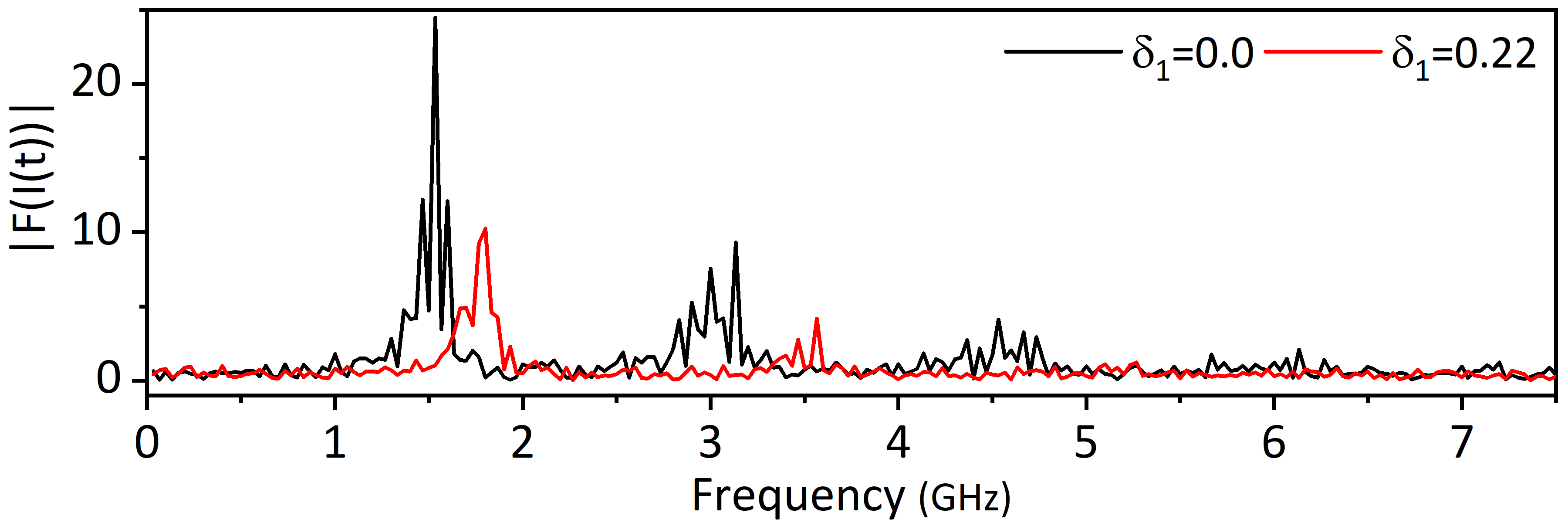}
	\caption{Power spectrum of oscillations of the maximum intensity point in a turbulent state obtained for $\mu=5.2$ and $r=2.4$. The dominant frequency of oscillations increases from 1.53 GHz to 1.80 GHz accompanied by a considerable drop in its amplitude when diffusion coefficient is increased from zero to 0.22.}
	\label{fft}
\end{figure}
\begin{figure*}
	\includegraphics[width=0.6\columnwidth]{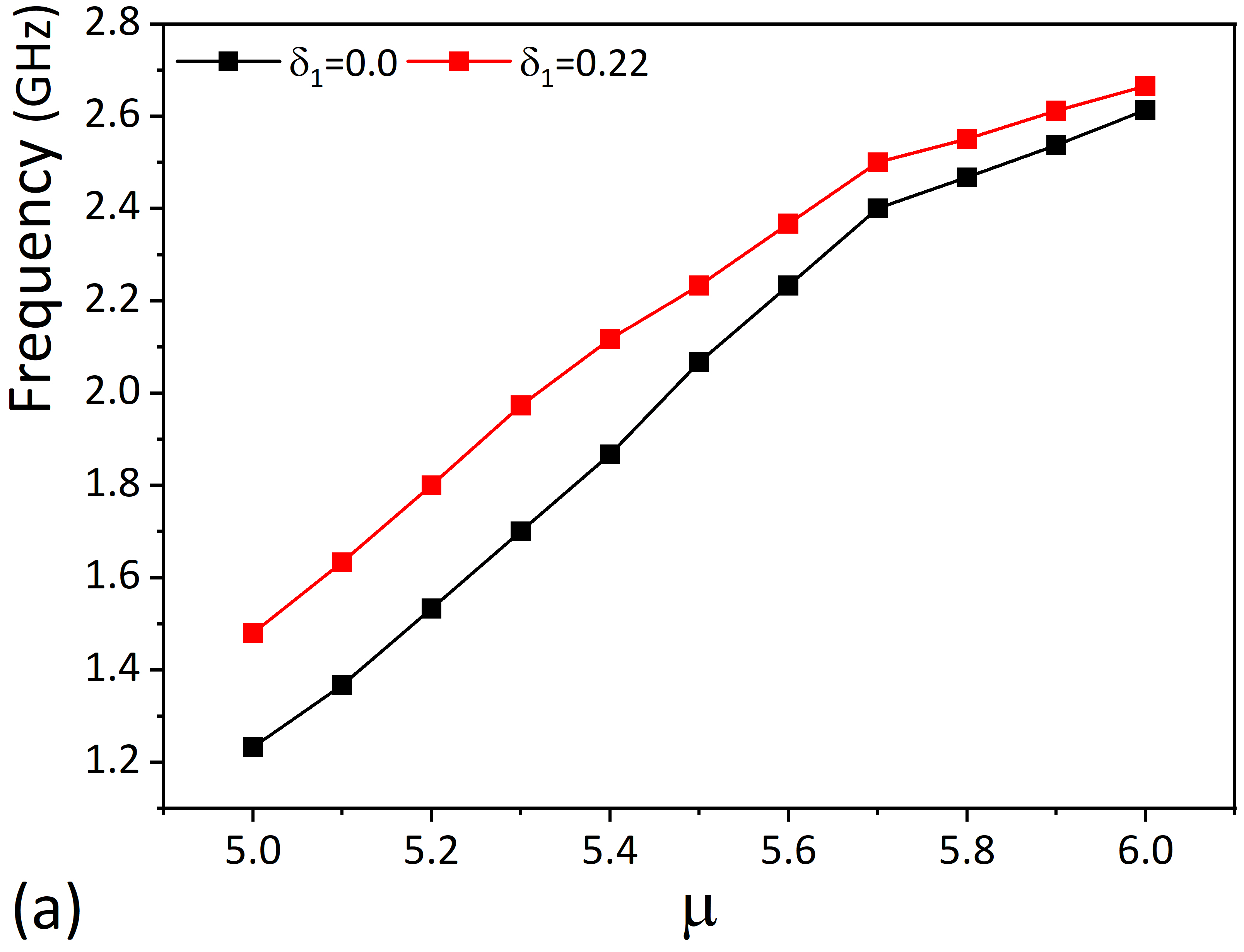}\quad \includegraphics[width=0.6\columnwidth]{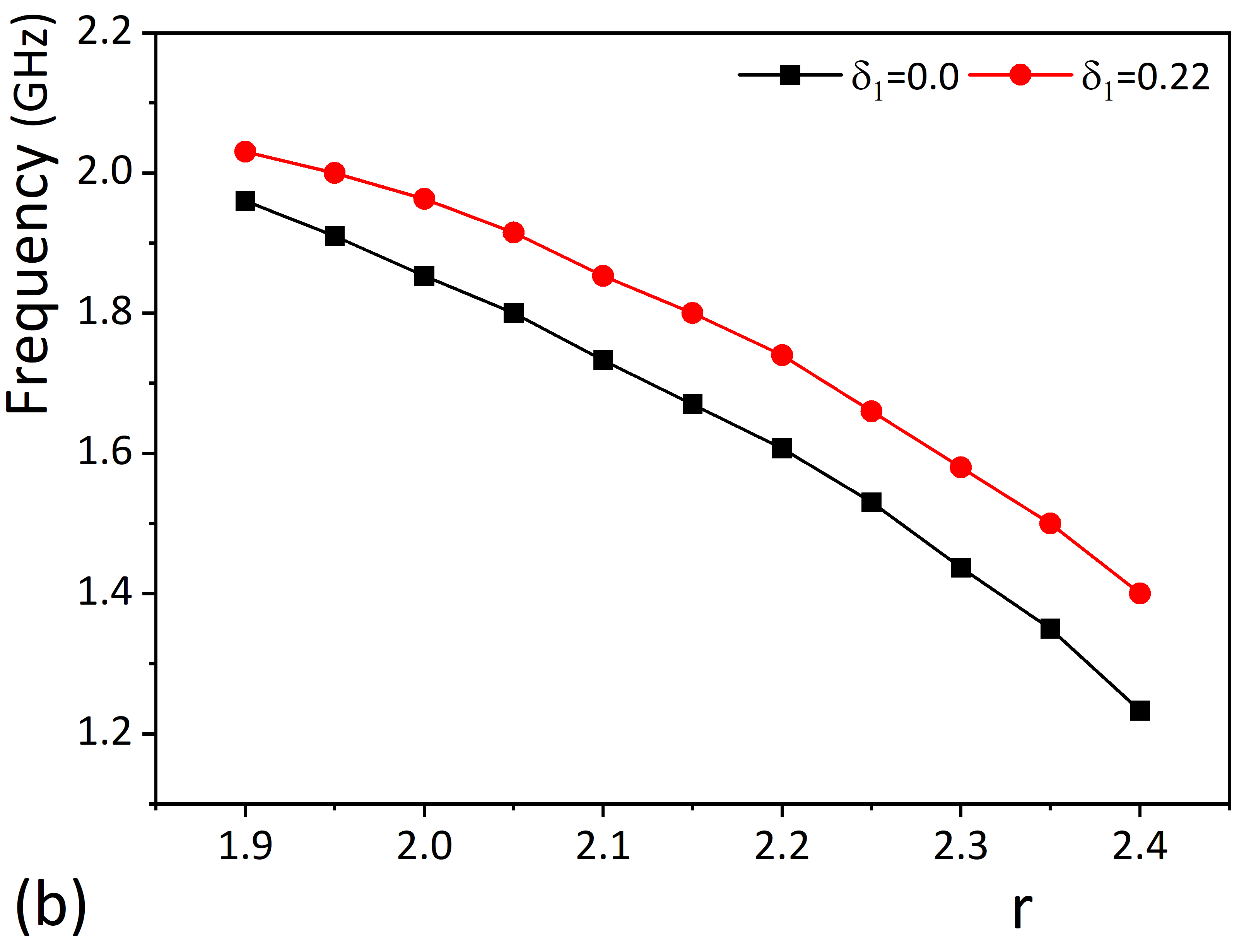}\quad \includegraphics[width=0.6\columnwidth]{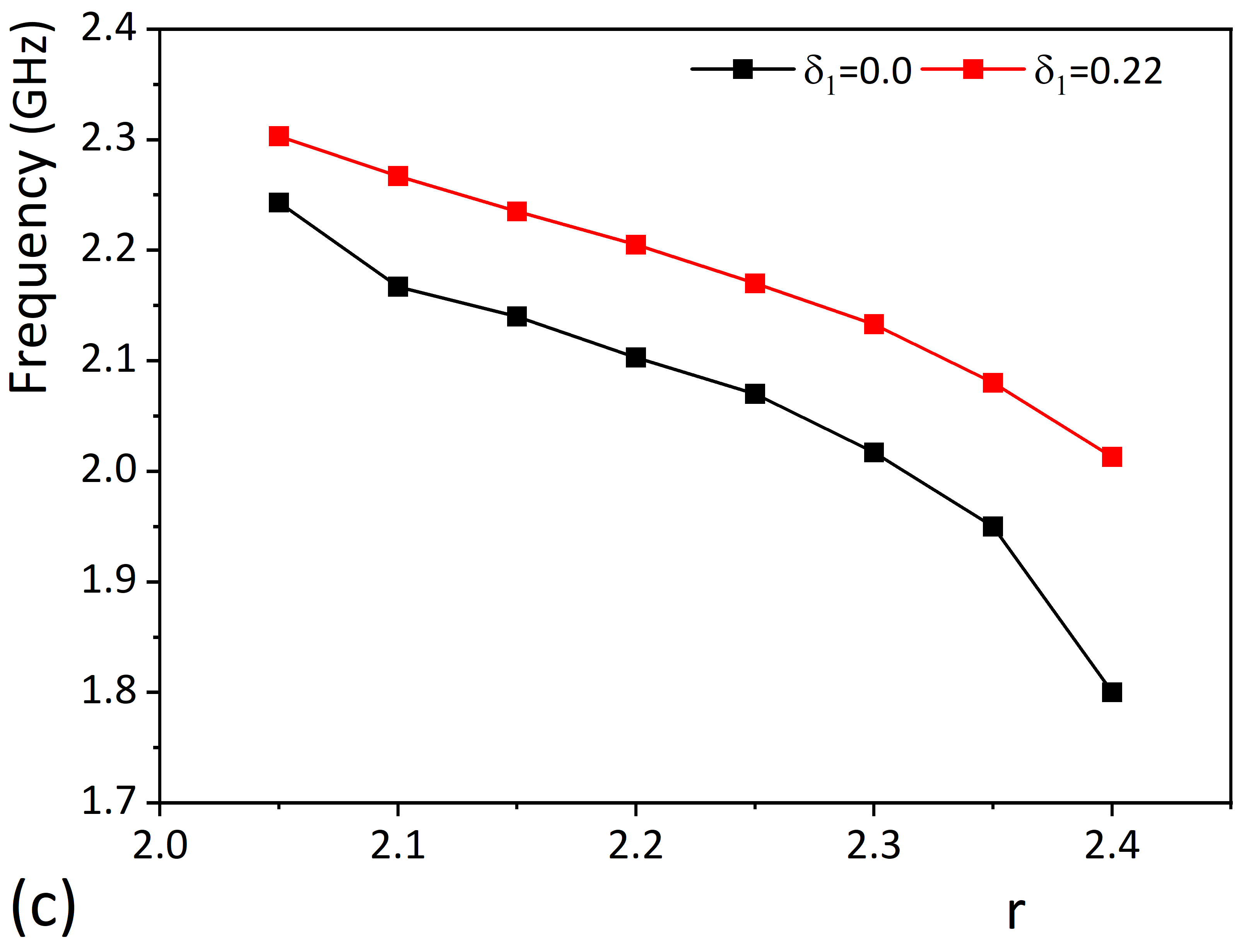}
	\caption{Dominant frequency of intensity oscillations versus $\mu$ compared for $\delta_1=0$ and $\delta_1=0.22$ when $r$ is fixed at 2.4 (a). The same versus $r$ for $\mu=5.0$ (b) and $\mu=5.35$ (c).}
	\label{freqmur}
\end{figure*}
\section{Model equations: inclusion of lateral carrier diffusion}
The effective model appropriate for RW studies in a broad-area semiconductor laser with an intracavity saturable absorber is basically the one used in \cite{CSL07} for investigation of CSs but with an additional term representing diffusion coefficient for the electric field $\delta_F$ that sets a filter to the number of spatial frequencies involved in the dynamics \cite{Rimoldi17,Eslami20}. While the diffusion term in the electric field equation makes the model completely suitable for studying spatiotemporal turbulence in terms of structural stability, we shall include lateral carrier diffusion in the equations for population dynamics in the active and passive media. The diffusion length $l$ is, by definition, the average distance that the excess carriers can cover before they recombine and depends on the lifetime and mobility of the carriers scaling as $\sqrt{C\tau}$ where $C$ is the diffusion constant and $\tau$ is the carrier lifetime before recombination. In our case of a semiconductor laser with a saturable absorber, carrier lifetimes are different in the two amplifying and absorbing medium which lead to a different diffusion coefficient for carriers. Then, we will have the following calculations for the diffusion lengths in the active and passive materials respectively:
\begin{equation}
	l_1=\sqrt{C\tau_c}, \quad l_2=\sqrt{C\tau_c/r}, 
\end{equation}
where $\tau_c$ is the carrier lifetime in the active material and $r$ is the ratio of the carrier lifetime in the amplifier to that in the absorber. In the model equations, the transverse diffusion coefficients for carriers in the amplifier and absorber are then defined respectively as
\begin{align}
	\delta_1&=l_1^2/\Bar{a}=C\tau_c/\Bar{a},\\ \nonumber
	\delta_2&=l_2^2/\Bar{a}=C\tau_c/r\Bar{a}=\delta_1/r, 
\end{align}
where $\Bar{a}$ is the diffraction coefficient. We scan $\delta_1$ from zero (no lateral diffusion at all) to 0.22 which translates to diffusion lengths in the order of zero to a few microns. With these arrangements, the full set of equations used in our simulations is written as
\begin{eqnarray}
	\nonumber
	\partial_t F&=&[(1-i\alpha)D+(1-i\beta)d-1+(\delta_F+i)\nabla_{\bot}^2]F\,,\\
	\nonumber
	\partial_t D&=&b[\mu-D(1+|F|^2)-BD^2+\delta_1 \nabla_{\bot}^2 D]\,,\\
	\partial_t d&=&rb[-\gamma-d(1+s|F|^2)-Bd^2+\dfrac{\delta_1}{r} \nabla_{\bot}^2 d]\,,
	\label{model}
\end{eqnarray}
where $F$ is the slowly varying amplitude of the electric field, and $D$, $d$ are the population variables defined as
\begin{equation}
	D=\eta_1(N_1/N_{1,0}-1),\; d=\eta_2(N_2/N_{2,0}-1).
	\label{popeqs}
\end{equation}
In Eq.~\ref{popeqs}, $N_1$  and $N_2$  are the carrier densities in the active and passive materials, respectively; $N_{1,0}$ and $N_{2,0}$ are their transparency values, and $\eta_1$, $\eta_2$ are dimensionless coefficients related to gain and absorption, respectively. $\delta_F$ is the diffusion coefficient for the electric field that protects structural stability of the equations. The parameters $\alpha$ ($\beta$) and $b$ are the linewidth enhancement factor in the active (passive) material and the ratio of the photon lifetime to the carrier lifetime in the active material. $\mu$ is the pump parameter of the active material, $\gamma$ is the absorption parameter of the passive material, $s$ is the saturation parameter, and $B$ is the coefficient of radiative recombination. Time is scaled to the photon lifetime, and space to the diffraction length. Typically a time unit is a few ps and a space unit $\sim 4\, \mu$m. The integration of the dynamical equations is performed by a split-step method separating the time and space derivatives on a $256\times256$ (and occasionally $512\times512$) spatial grids with space step 0.25 implying the physical distance of 1 $\mu$m between two consecutive grid points. This integration method consists in separating the linear part (the Laplacian), integrated by a FFT algorithm, and the nonlinear part, integrated via a Runge-Kutta method. In all simulations, we use periodic boundary conditions unless stated otherwise. We have used the following parameter values throughout the paper: $\delta_F=0.01, s=1, B=0.1, \alpha=2, \beta=1, b=0.01$ and $\gamma=2$. $r$ and $\mu$ are used as the control parameters.\\
\begin{figure}
	\includegraphics[width=0.49\columnwidth]{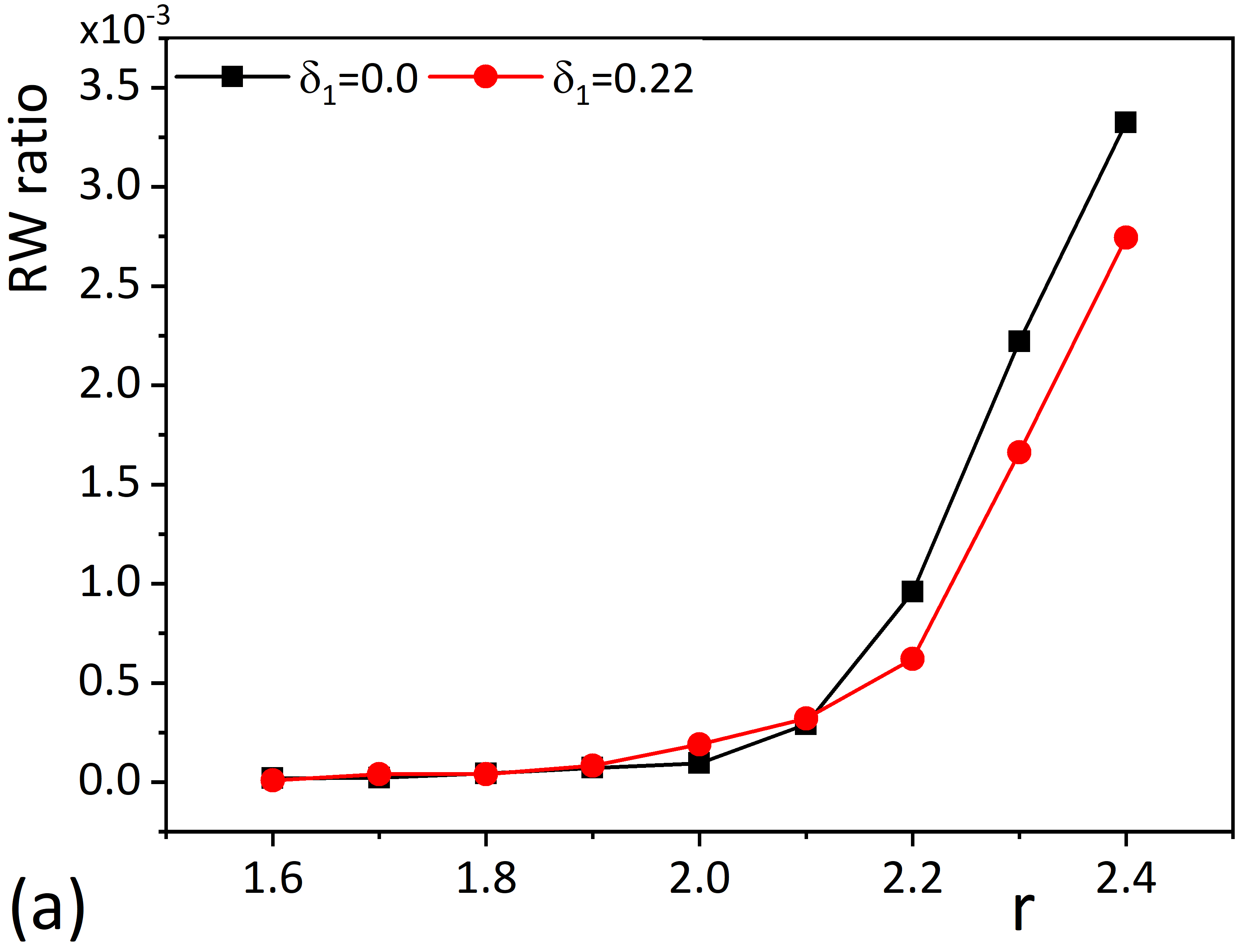} \includegraphics[width=0.49\columnwidth]{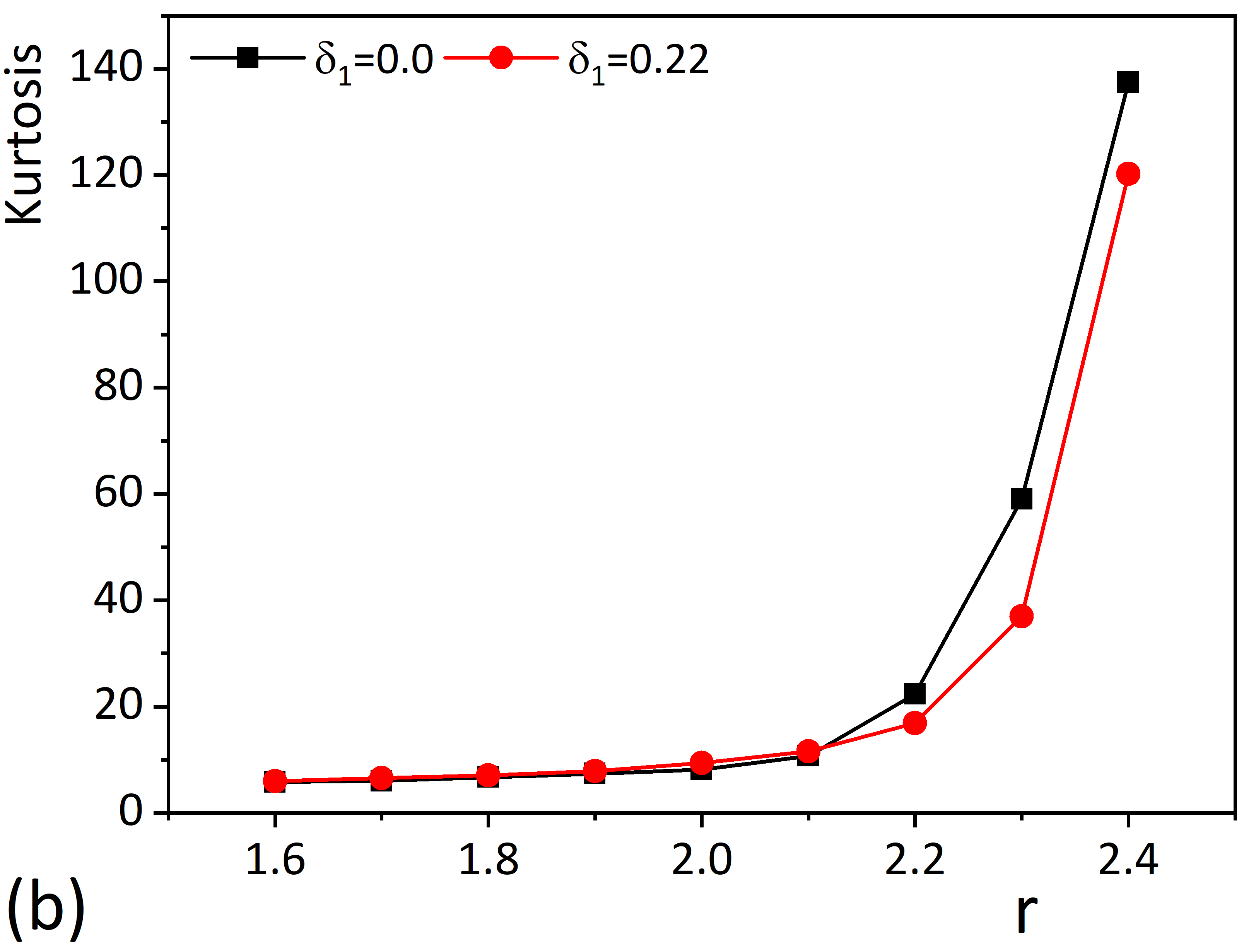}\\
	\includegraphics[width=0.49\columnwidth]{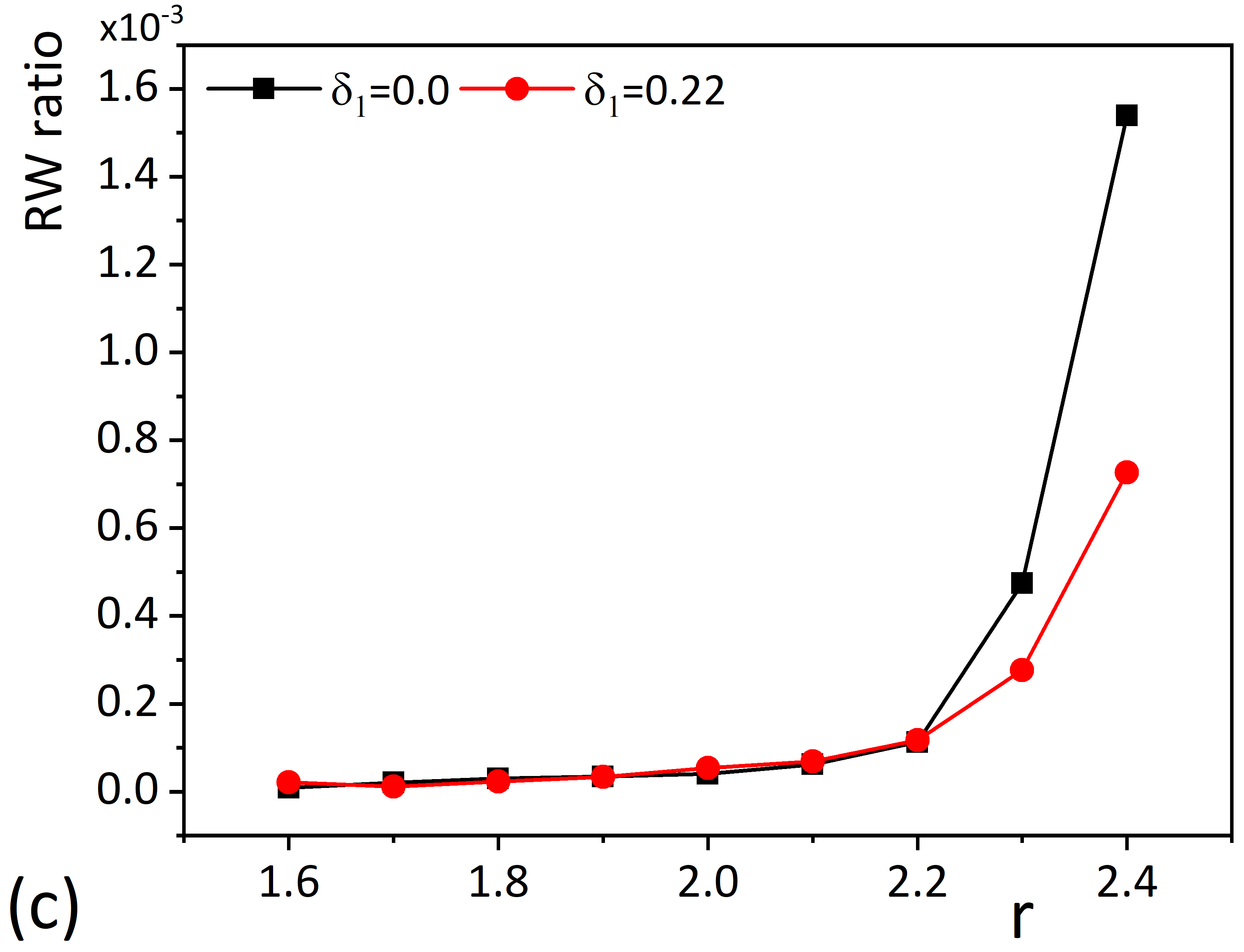} \includegraphics[width=0.49\columnwidth]{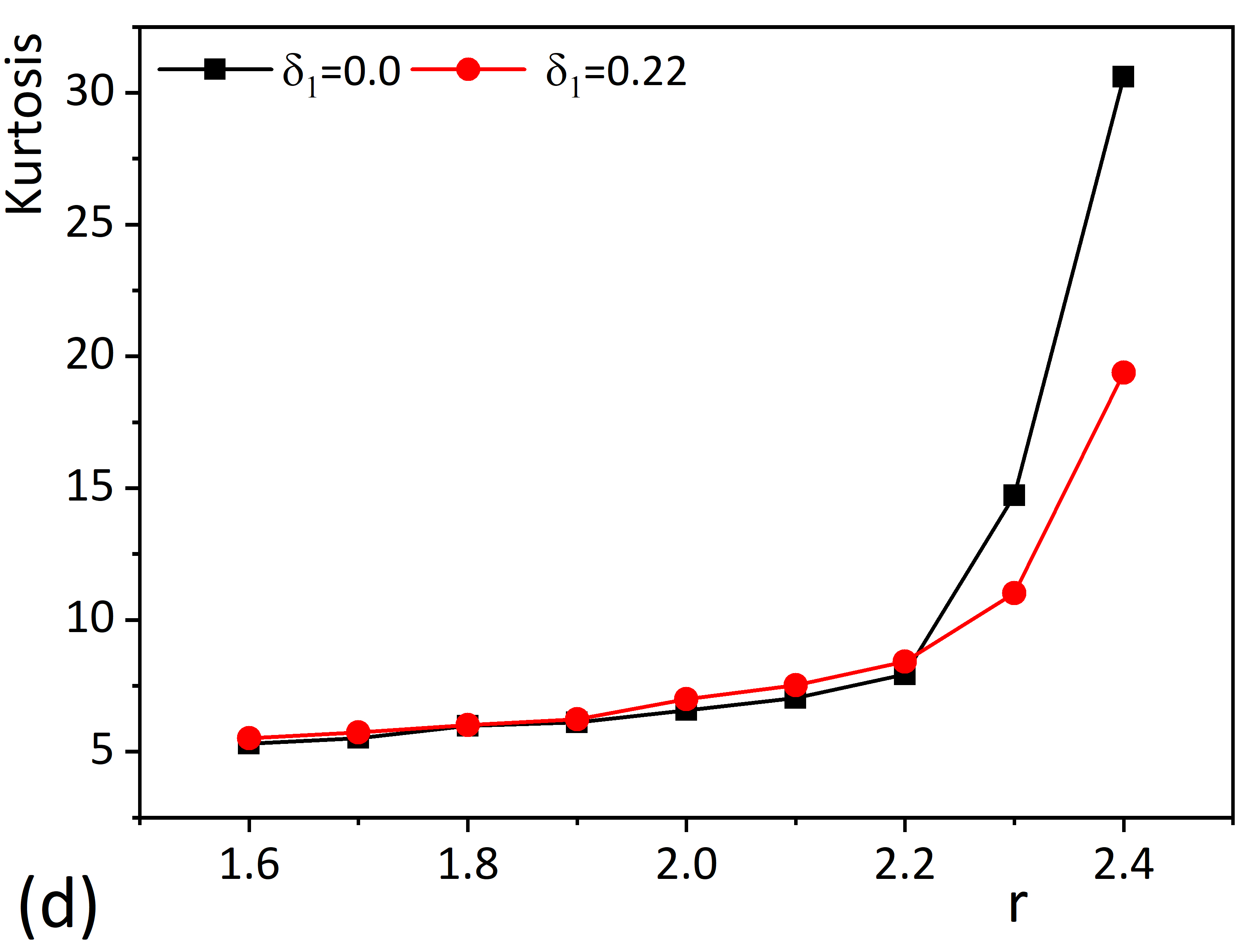}
	\caption{RW ratio (left panel) and kurtosis (right panel) compared for $\delta_1=0$ and $\delta_1=0.22$ in $r$ scan; below laser threshold $\mu=5.0$ (a,b) and above laser threshold $\mu=5.35$ (c,d).}
	\label{RWkbeab}
\end{figure}
\section{Turbulent solutions affected by transverse carrier diffusion}
In this section we present the effects of lateral carrier diffusion on the dynamics of turbulent solutions where RWs are likely to be emitted. In Fig.~\ref{branches} we only focus on the turbulent solutions along with the homogeneous steady state curve although, depending on parameters, there could be other branches of solutions such as stationary CSs, self-pulsing CSs and chaotic CSs which are detailed in \cite{Rimoldi17,Eslami20}. Fig.~\ref{branches} shows the solutions space where the maximum intensity of turbulent solutions averaged in time is compared in the two cases of $\delta_1=0$ and $\delta_1=0.22$. It is observed from Fig.~\ref{branches}(a) that the inclusion of transverse carrier diffusion has larger effects in lower pump current $\mu$ values which is consistent with Fig.~\ref{branches}(b) and (c) where turbulent branches are sketched in $r$ scan respectively for below $\mu=5.0$ and above $\mu=5.35$ laser threshold. We note that for $r=1$, where equal carrier lifetimes are considered for both amplifier and absorber materials, the time-averaged maximum intensity of turbulent states grows almost linearly with $\mu$, see \cite{Eslami20}.\\
\begin{figure}
	\includegraphics[width=0.49\columnwidth]{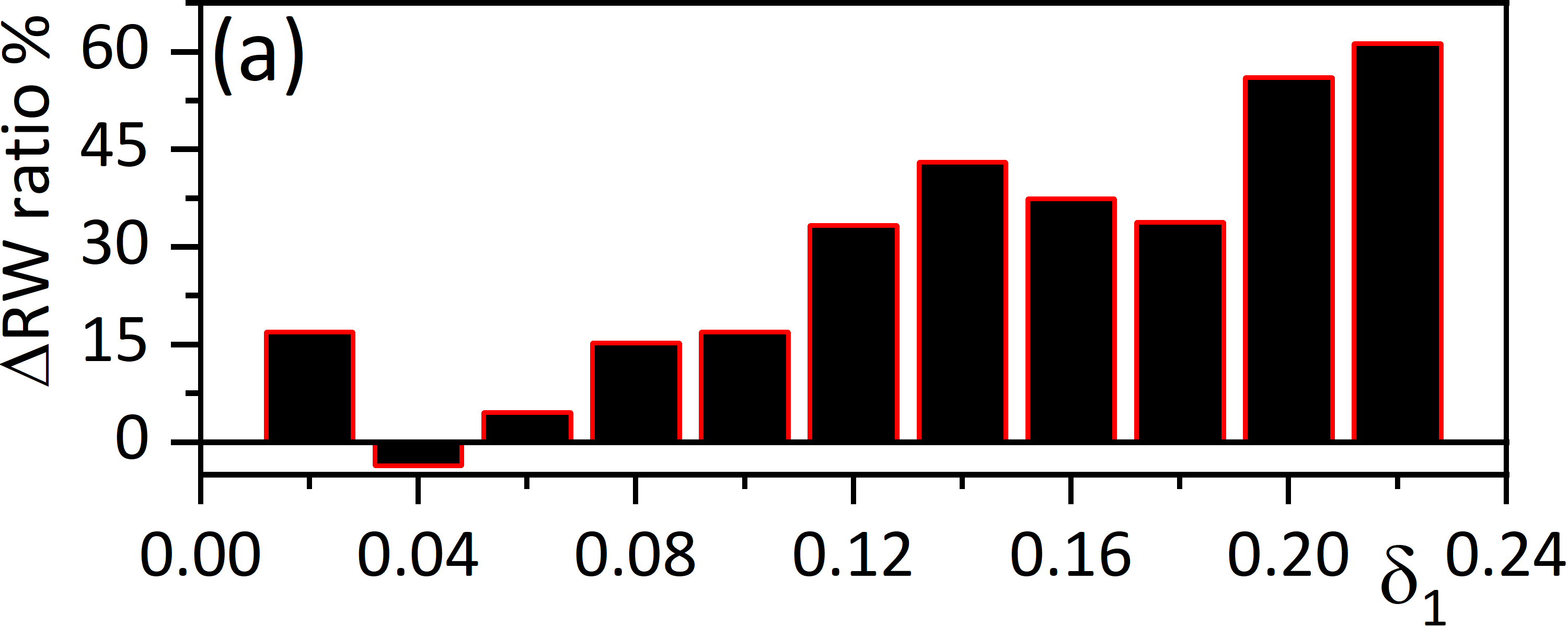} \includegraphics[width=0.49\columnwidth]{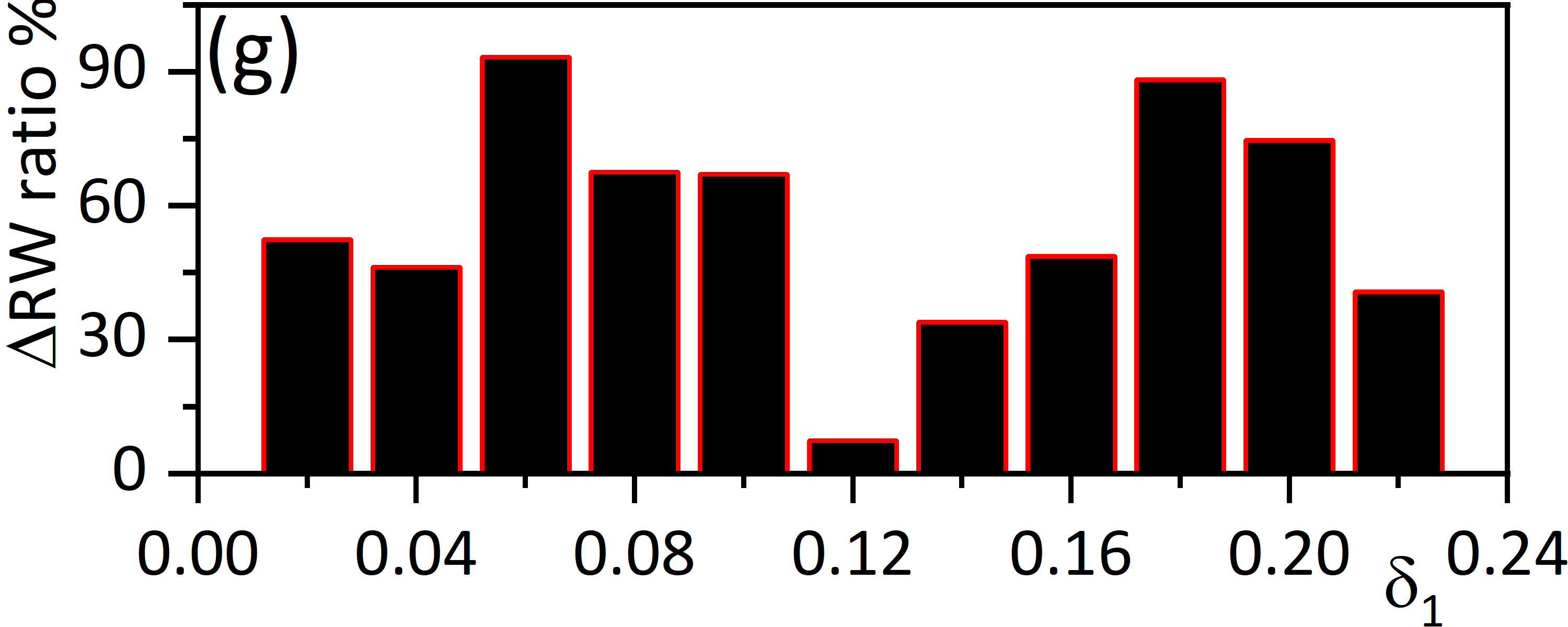}\\
	\includegraphics[width=0.49\columnwidth]{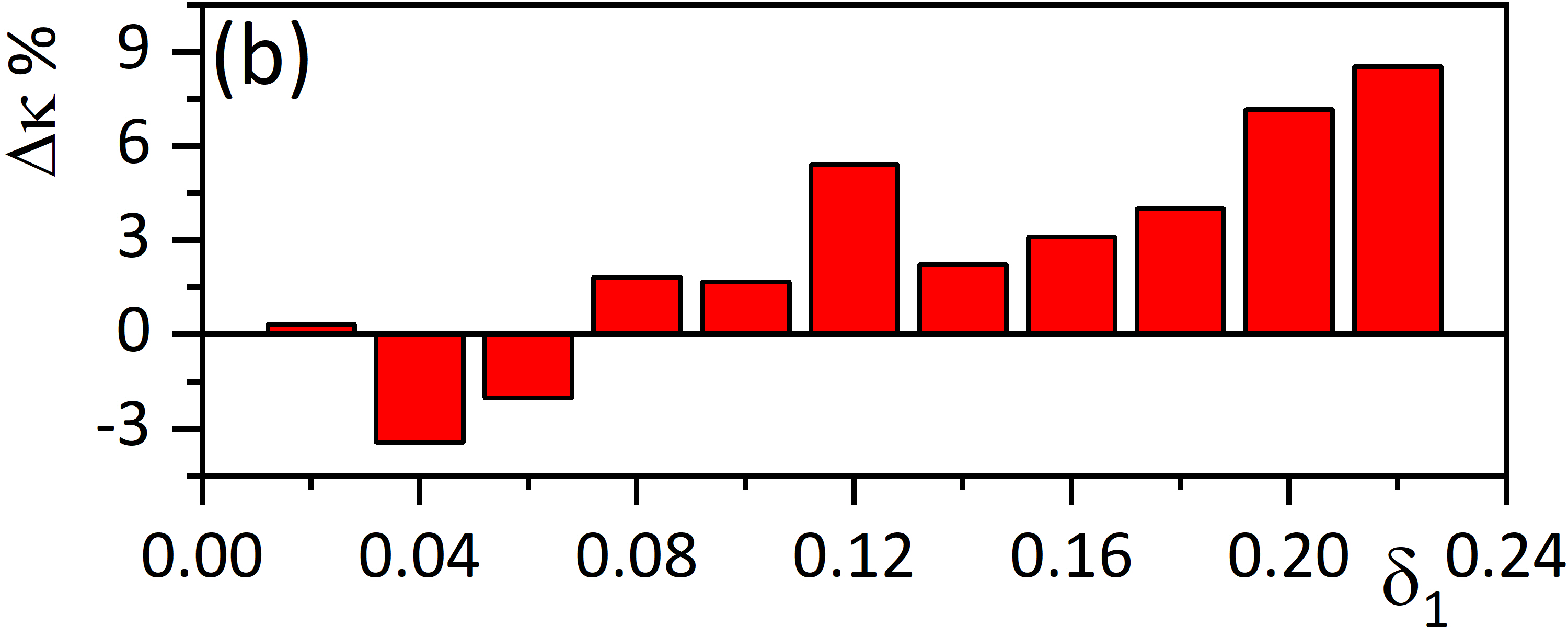} \includegraphics[width=0.49\columnwidth]{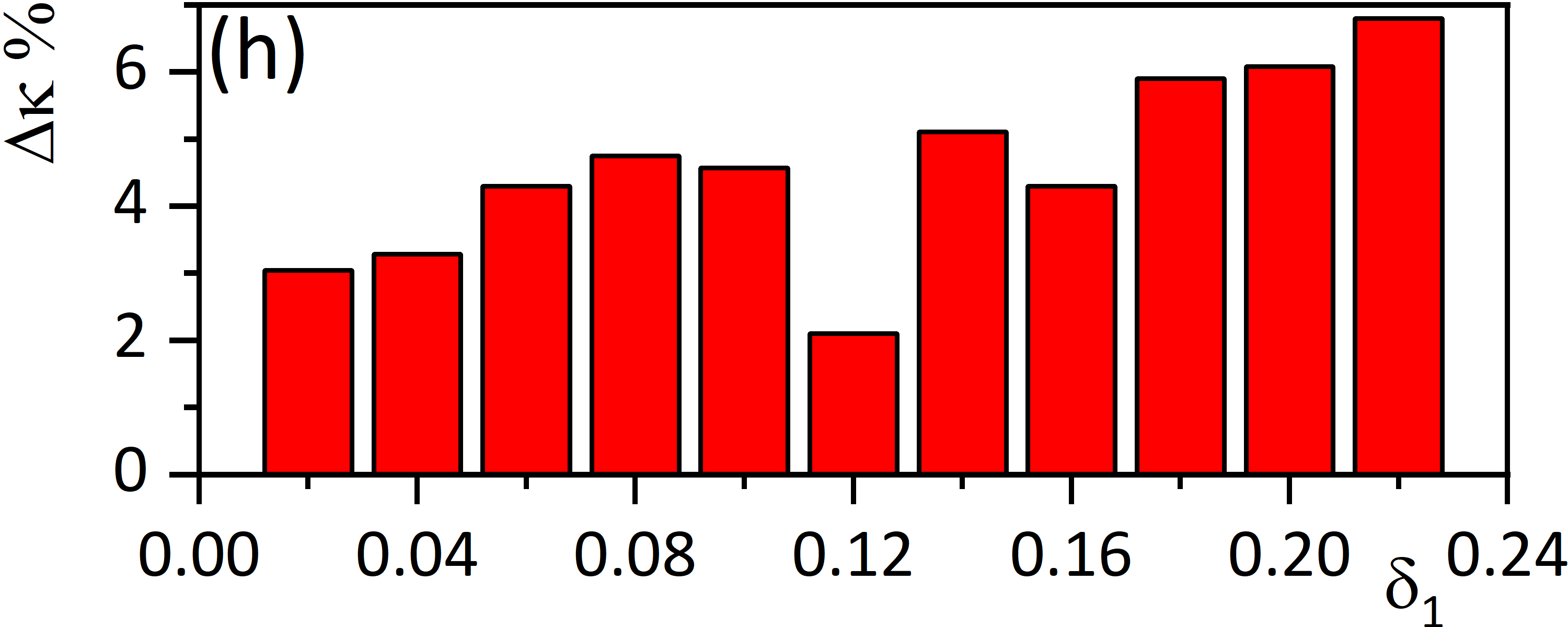}\\
	\includegraphics[width=0.49\columnwidth]{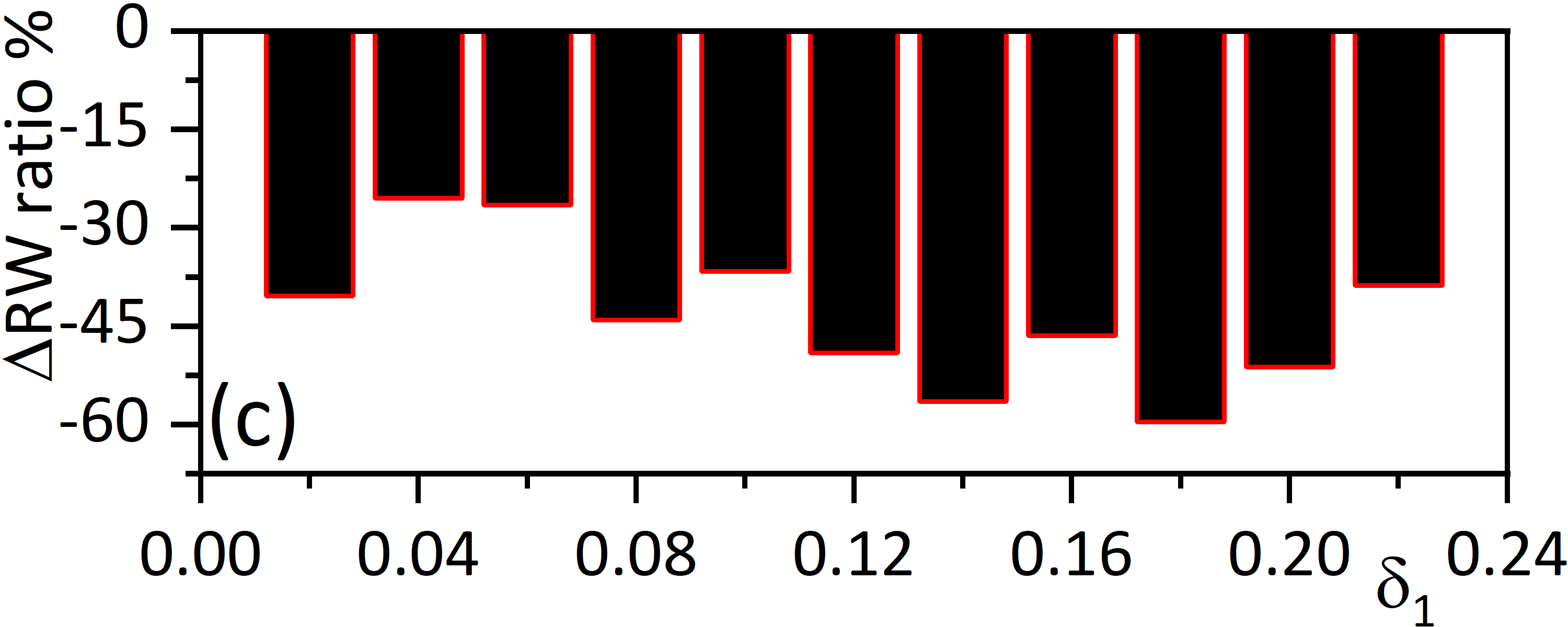} \includegraphics[width=0.49\columnwidth]{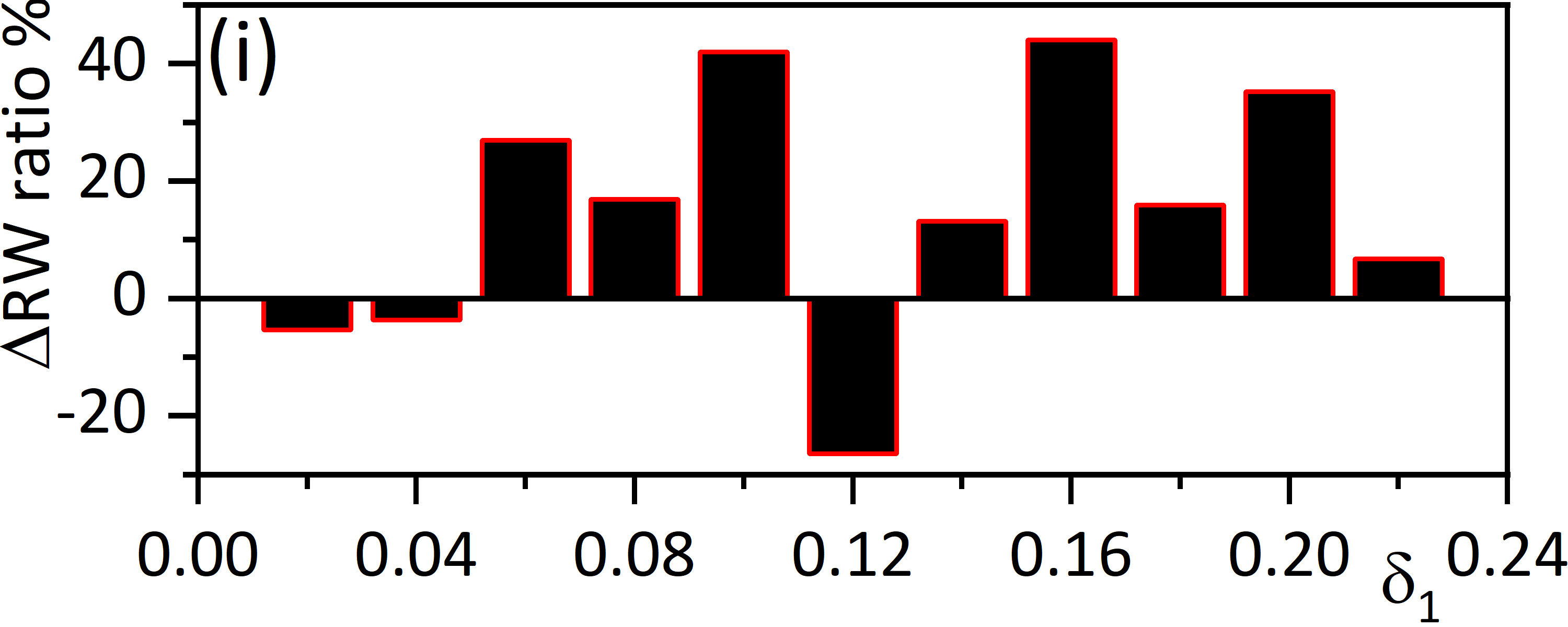}\\
	\includegraphics[width=0.49\columnwidth]{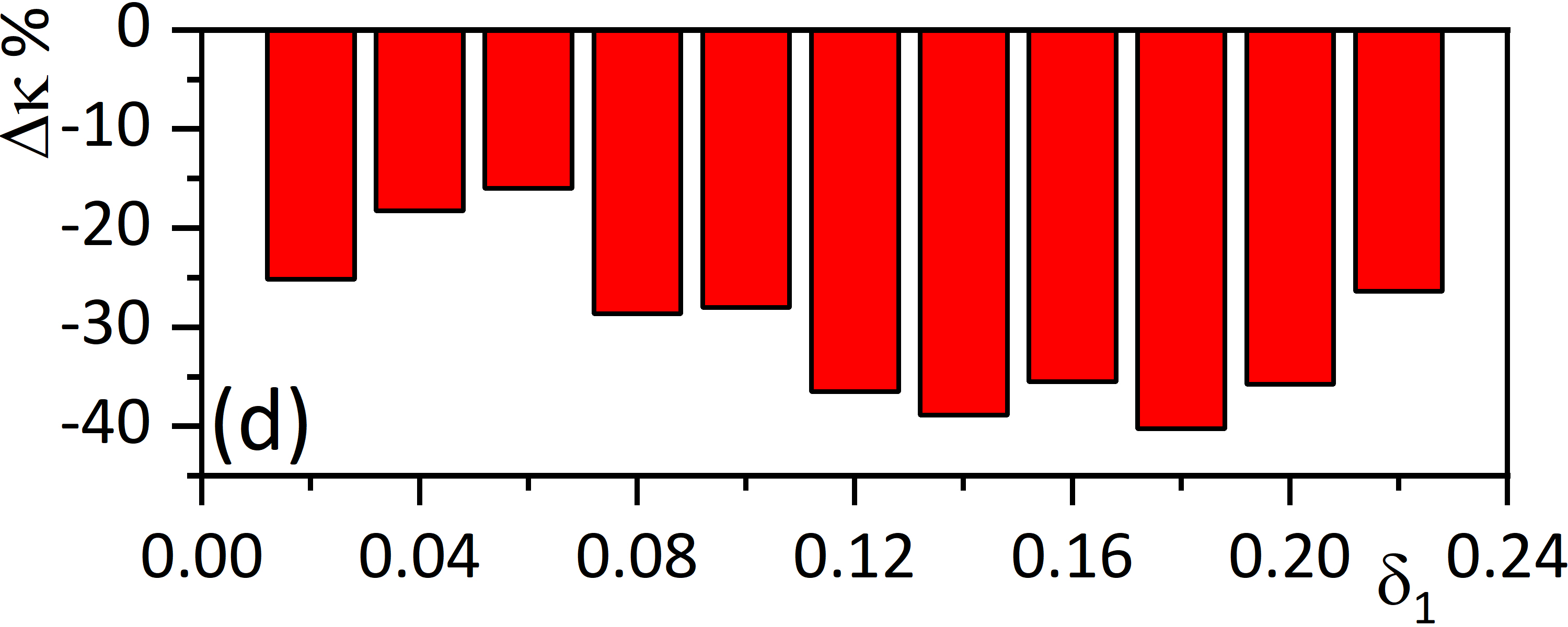} \includegraphics[width=0.49\columnwidth]{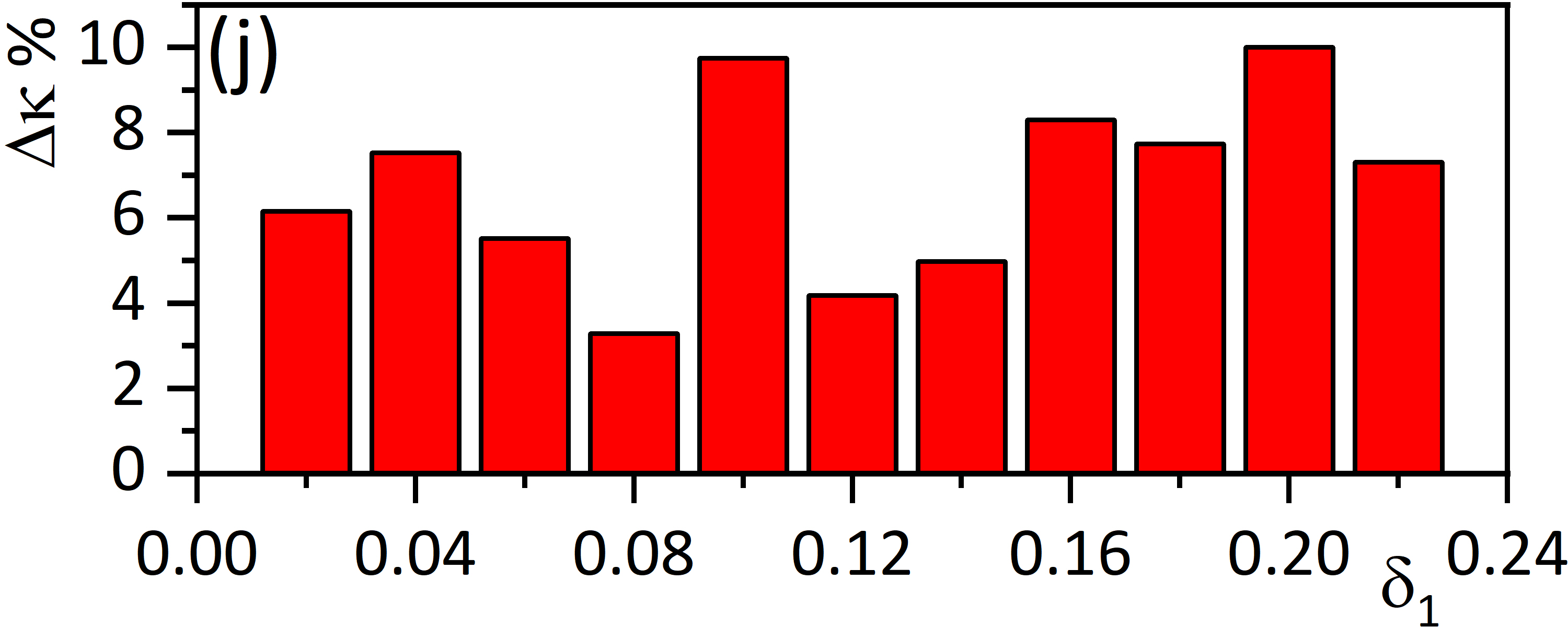}\\
	\includegraphics[width=0.49\columnwidth]{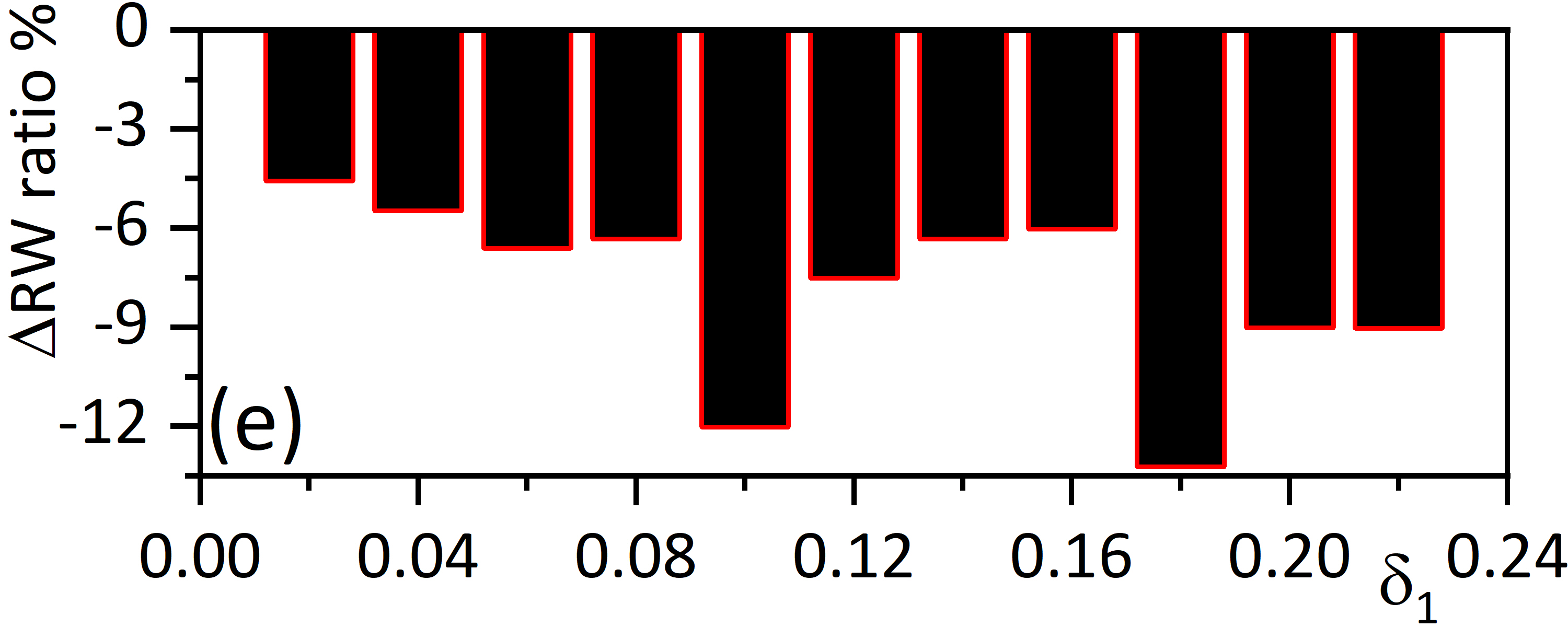} \includegraphics[width=0.49\columnwidth]{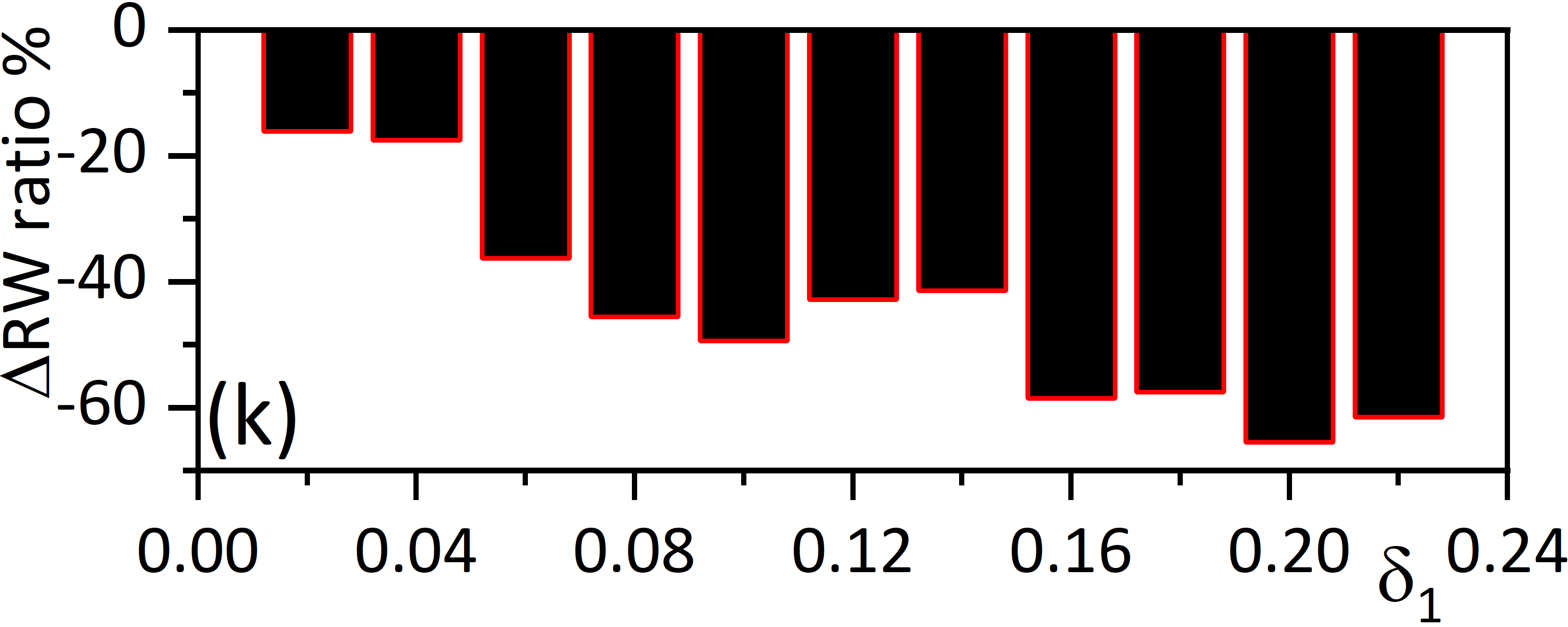}\\
	\includegraphics[width=0.49\columnwidth]{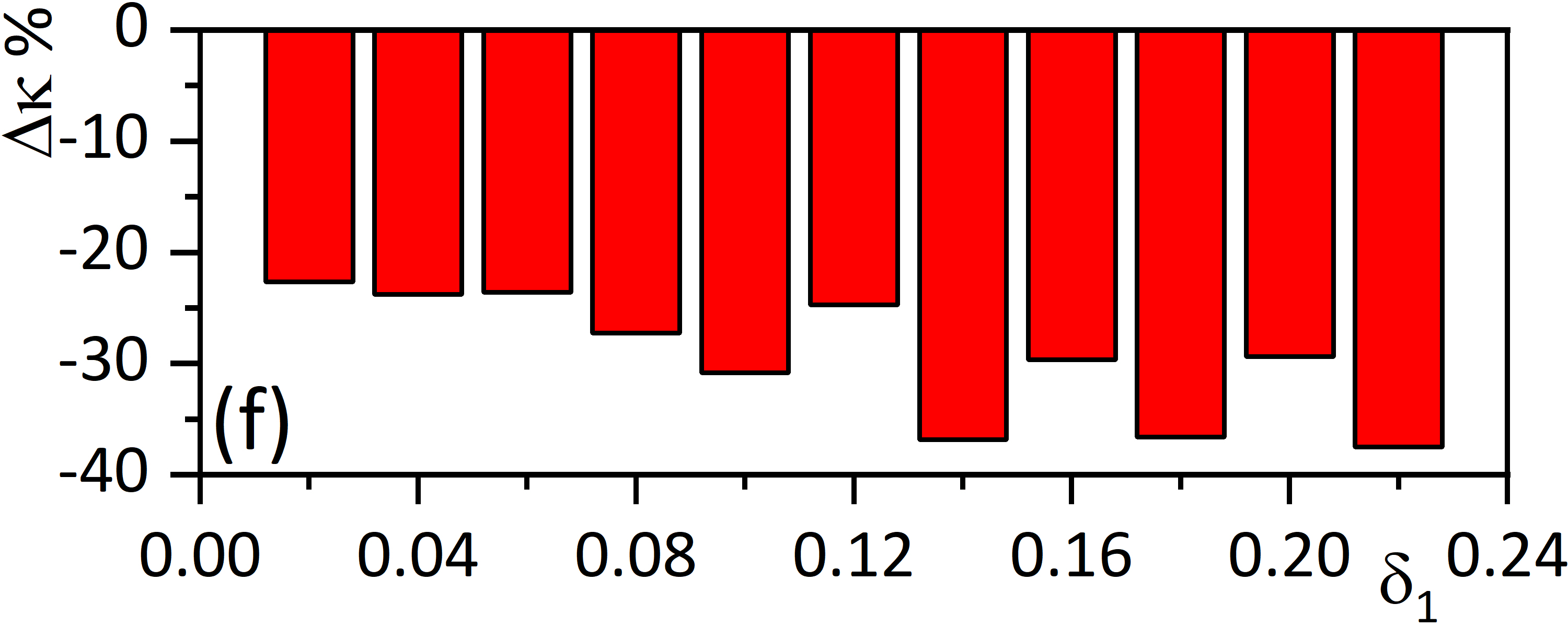} \includegraphics[width=0.49\columnwidth]{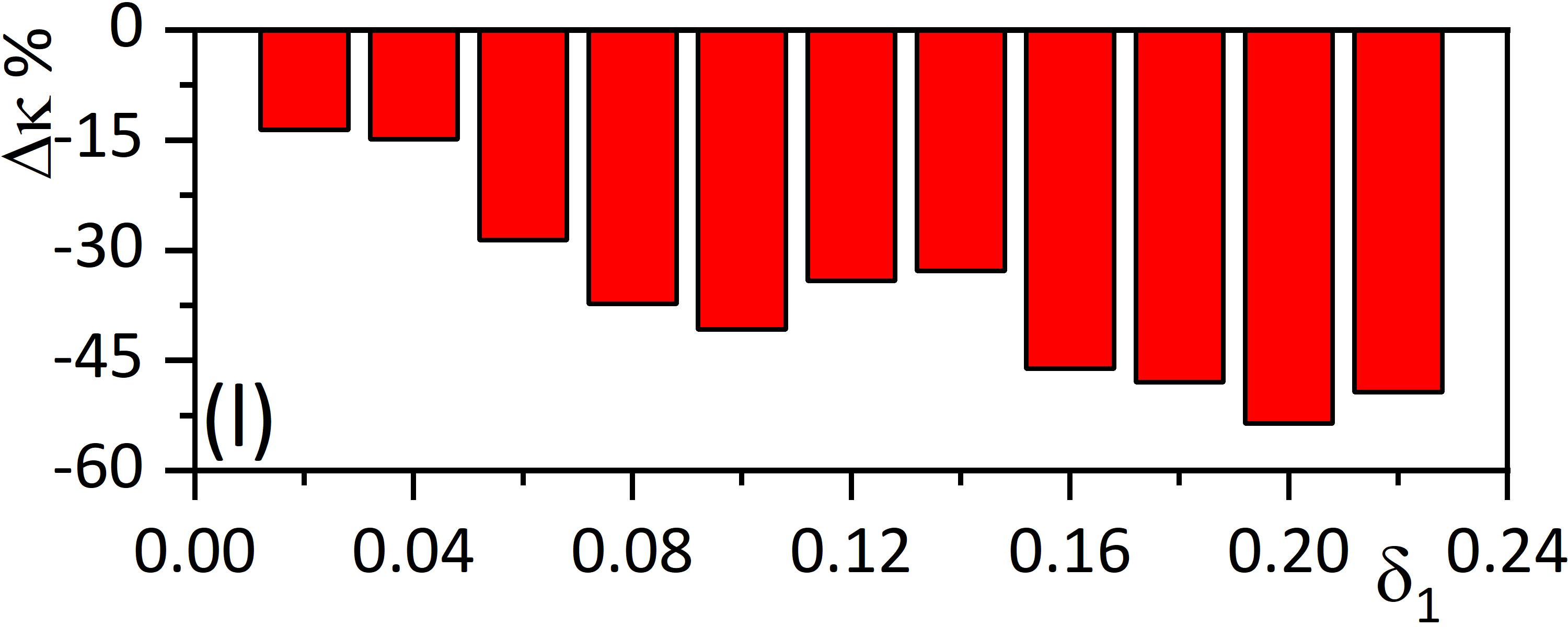}\\
	\caption{RW ratio and kurtosis difference from $\delta_1=0$ in percentage versus $\delta_1$. $r=2.0$ (a,b) and (g,h), $r=2.2$ (c,d) and (i,j), and $r=2.4$ (e,f) and (k,l). The left panel shows the pump regime of below laser threshold ($\mu=5.0$) and the right panel shows that of above laser threshold ($\mu=5.35$). We remind that for $r=2.4$ larger diffusion coefficient than $\delta_1=0.22$ switches the laser to the off state.}
	\label{RWkdelta}
\end{figure}
Transverse diffusion of carriers can also have consequences in terms of the dominant frequencies of intensity oscillations in a turbulent state as it allows drift of carriers within a finite area before recombination to emit a photon. We have compared the situation with $\delta_1=0$ to that with $\delta_1=0.22$ for a typical case of $\mu=5.2$ and $r=2.4$ in Fig.~\ref{fft} which clearly shows an increase in the dominant frequency and a reduction in its amplitude. The trends of variations in the dominant frequency of intensity oscillations in $\mu$ and $r$ scans are shown in Fig.~\ref{freqmur} for the two values of $\delta_1=0$ and $\delta_1=0.22$. It is observed from Fig.~\ref{freqmur}(a) that increase in the dominant frequencies persists for the entire range of $\mu$ values for nonzero lateral diffusion coefficient in spite of the fact that for higher carrier densities provided by large pump currents the difference from $\delta_1=0$ is small. The trend for dominant frequency of intensity oscillations is decreasing in $r$ scan while still maintaining its larger value compared to the case with zero diffusion at the same $r$, as depicted in Fig.~\ref{freqmur}(b) for pump current of below laser threshold and (c) for that above laser threshold.\\
\begin{figure}
	\includegraphics[width=0.49\columnwidth]{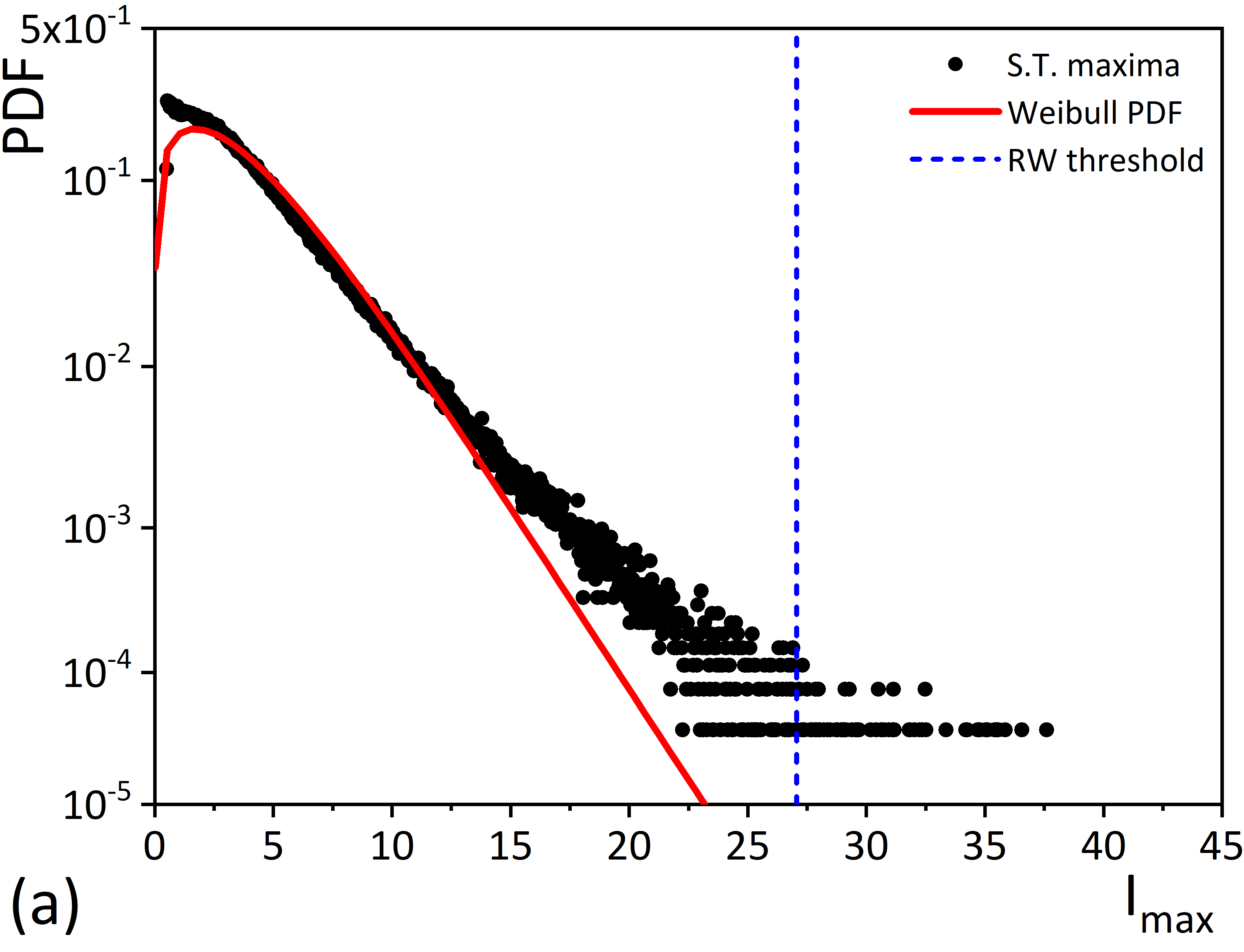} \includegraphics[width=0.49\columnwidth]{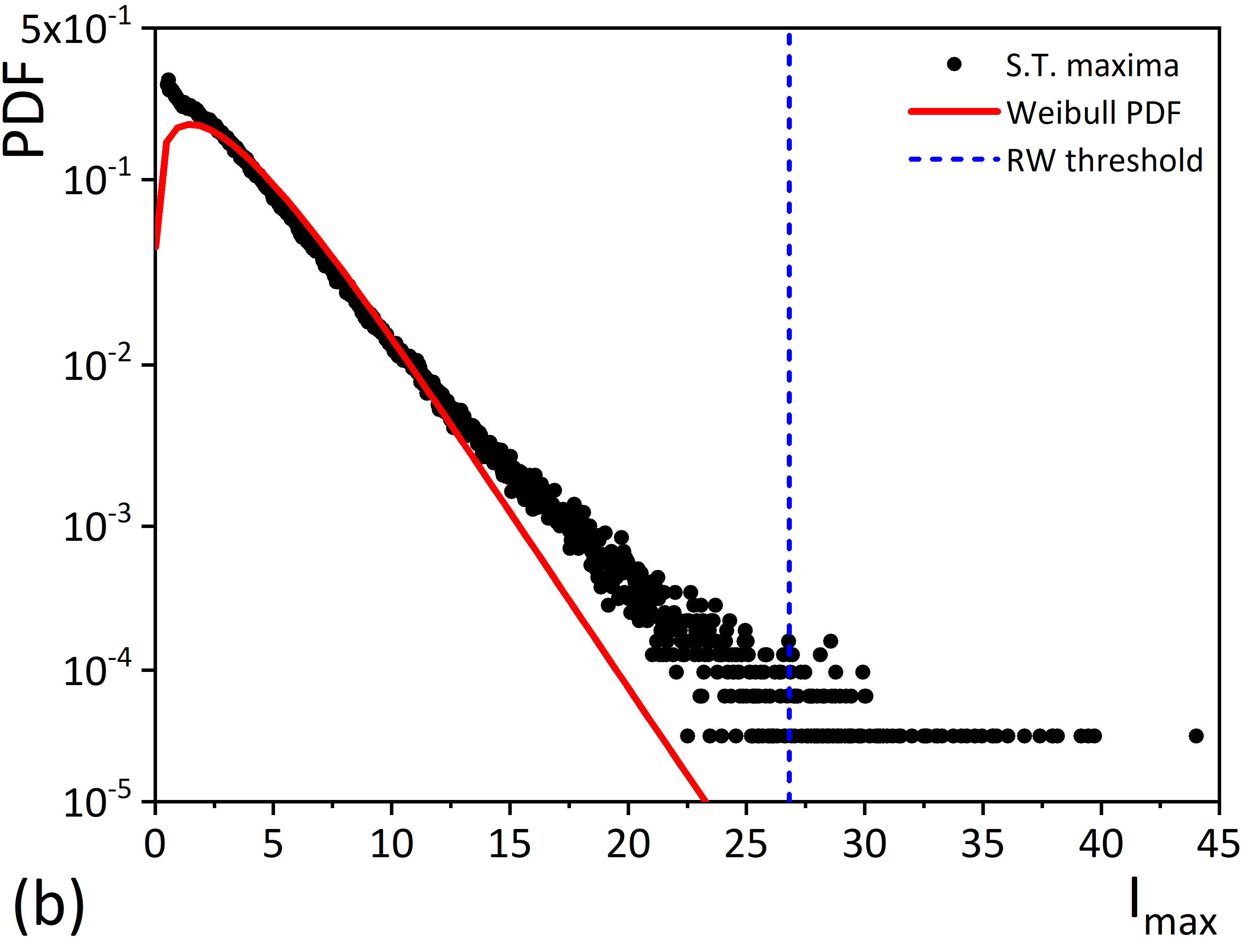}
	\caption{PDFs calculated over spatiotemporal intensity maxima and the corresponding RW thresholds. Weibull fit is used to distinguish those spatiotemporal events which deviate from the normal. Number of RWs increases from 76 for $\delta_1=0$ (a) to 130 for $\delta_1=0.22$ (b) in a simulation window of 50 ns. Other values are $\mu=5.0$ and $r=2.0$.}
	\label{PDF2.0}
\end{figure}
\begin{figure}
	\includegraphics[width=0.49\columnwidth]{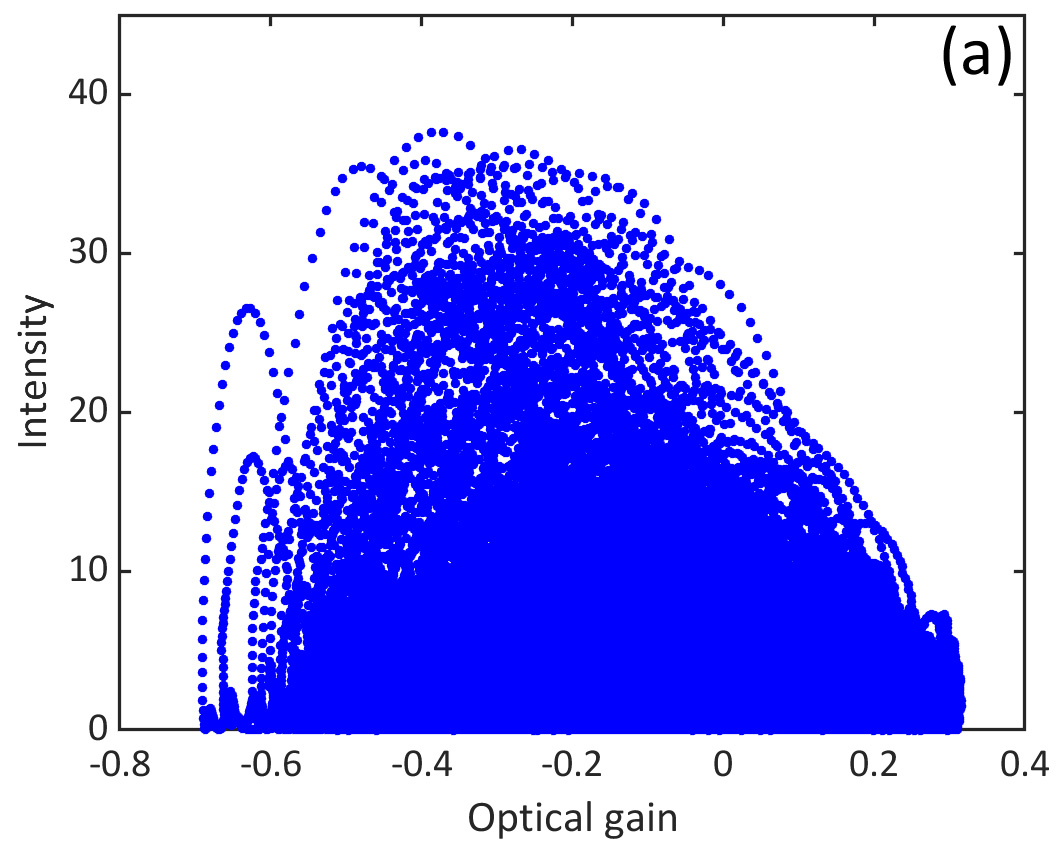} \includegraphics[width=0.49\columnwidth]{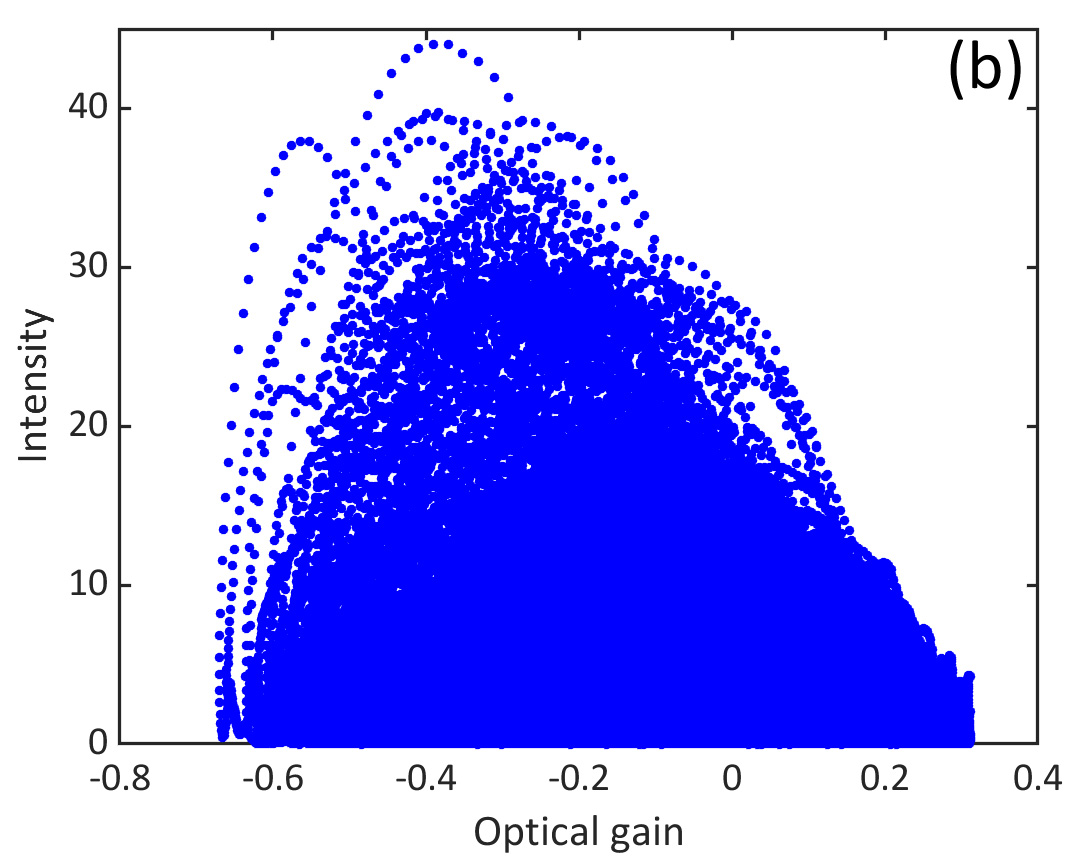}\quad
	\caption{Trajectories of spatiotemporal maxima in sub-space made up of optical gain (D+d-1) and intensity for $\delta_1=0$ (a) and $\delta_1=0.22$ (b). RW ratio increases in (b) where the area relevant to extreme intensities is more touched by the trajectories. Other values are $\mu=5.0$ and $r=2.0$.}
	\label{gainInt2.0}
\end{figure}
\section{Rogue waves affected by transverse carrier diffusion}
RWs in this system of two dimensions are generated in the peaks of a turbulent state and satisfy the widely-accepted threshold defined for characterizing RWs, i.e. $\langle I \rangle + 8 \sigma$, where $\sigma$ is the standard deviation. However, we take a rather stringent threshold condition for RWs which uses the probability density function (PDF) of spatiotemporal maxima instead of total intensity and distinguishes the RWs among many spatiotemporal maxima according to $I_{RW}=\langle I_{max} \rangle + 8 \sigma$ \cite{Rimoldi17}. In this case, the Weibull distribution
\begin{equation}
	\dfrac{a}{b}\left(\dfrac{I_{max}}{b}\right)^{a-1}\exp{\left[-\left(\dfrac{I_{max}}{b}\right)^a\right]},
\end{equation}
is a better choice to evidence the deviations leading to characterizing a wave as rogue wave \cite{Rimoldi17,Eslami20}. We also make use of two other indicators to measure the degree of rogueness, RW ratio and kurtosis respectively corresponding to the number of spatiotemporal events with intensities exceeding the intensity threshold $I_{RW}$ to the total number of spatiotemporal events during simulation and the ratio of the fourth moment about the mean to the square of the variance given by
\begin{equation}
	\mathcal{K}=\dfrac{\dfrac{1}{n}\sum_{i=1}^n(I_i-\langle I \rangle )^4}{\left[\dfrac{1}{n}\sum_{i=1}^n (I_i-\langle I \rangle)^2 \right]^2}.
\end{equation}
\begin{figure}
	\includegraphics[width=0.49\columnwidth]{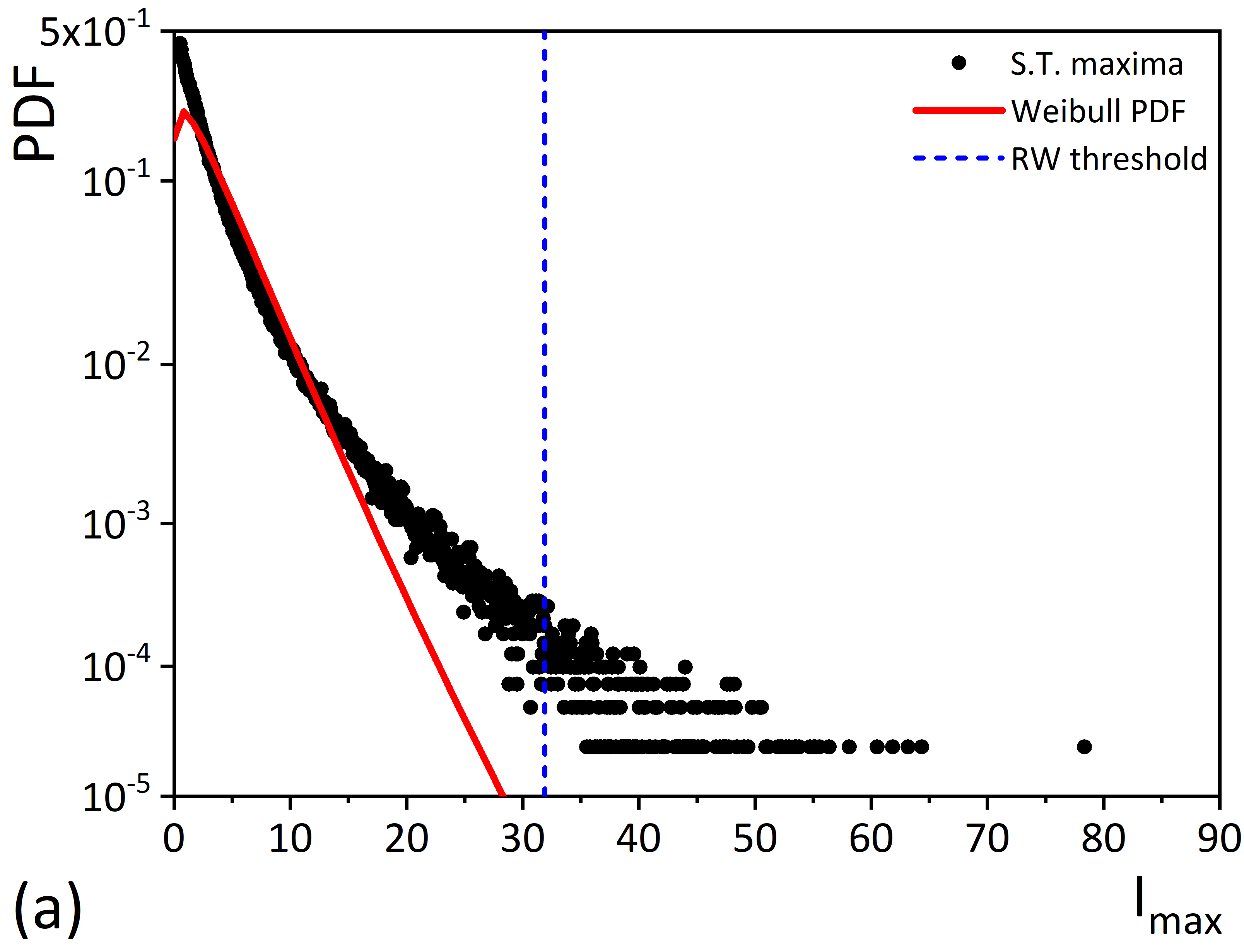} \includegraphics[width=0.49\columnwidth]{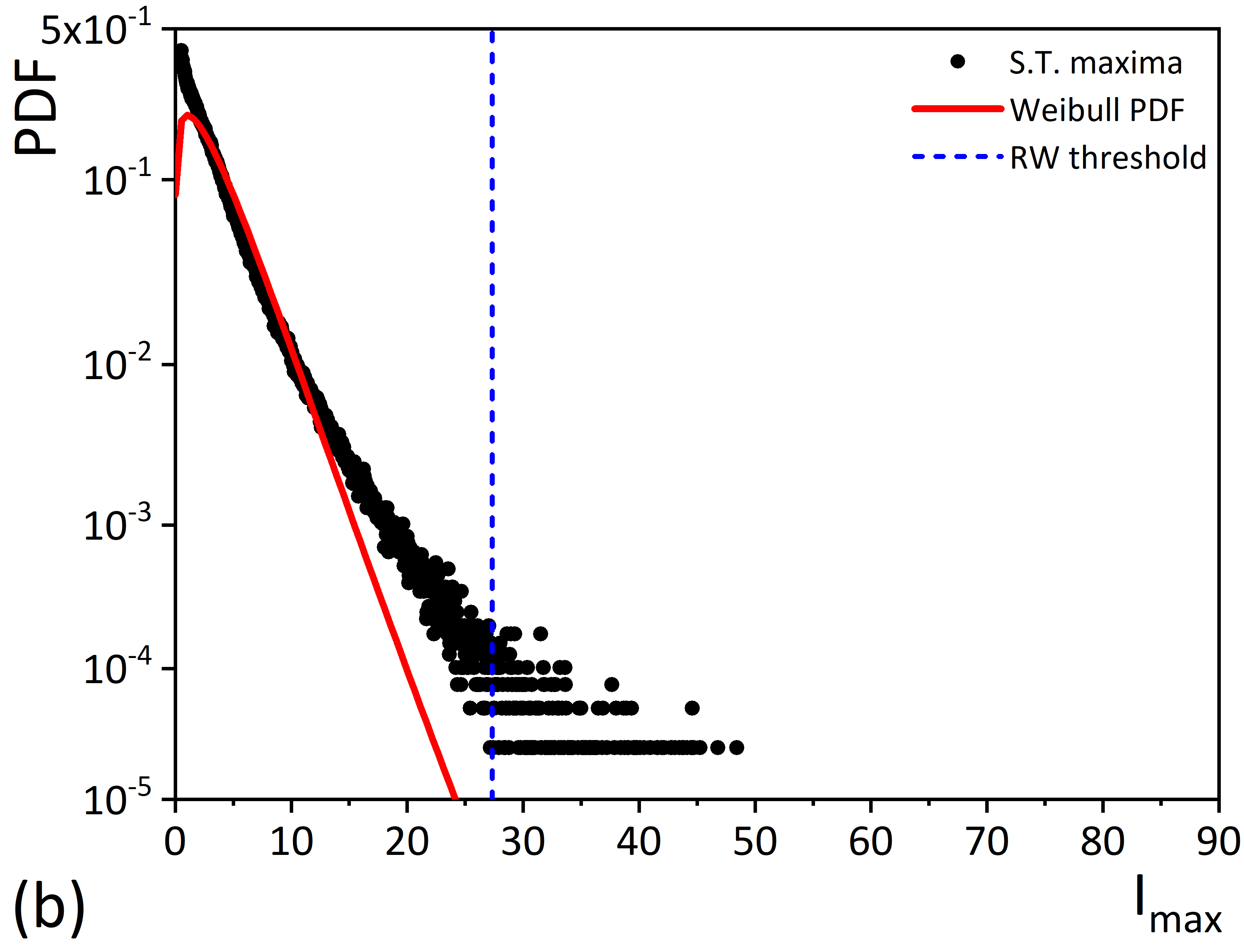}
	\caption{PDFs calculated over spatiotemporal intensity maxima and the corresponding RW thresholds. Weibull fit is used to distinguish those spatiotemporal events which deviate from the normal. Number of RWs decreases from 750 for $\delta_1=0$ (a) to 354 for $\delta_1=0.18$ (b) in the simulation window of 50 ns. Other values are $\mu=5.0$ and $r=2.2$.}
	\label{PDF2.2}
\end{figure}
\begin{figure}
	\includegraphics[width=0.49\columnwidth]{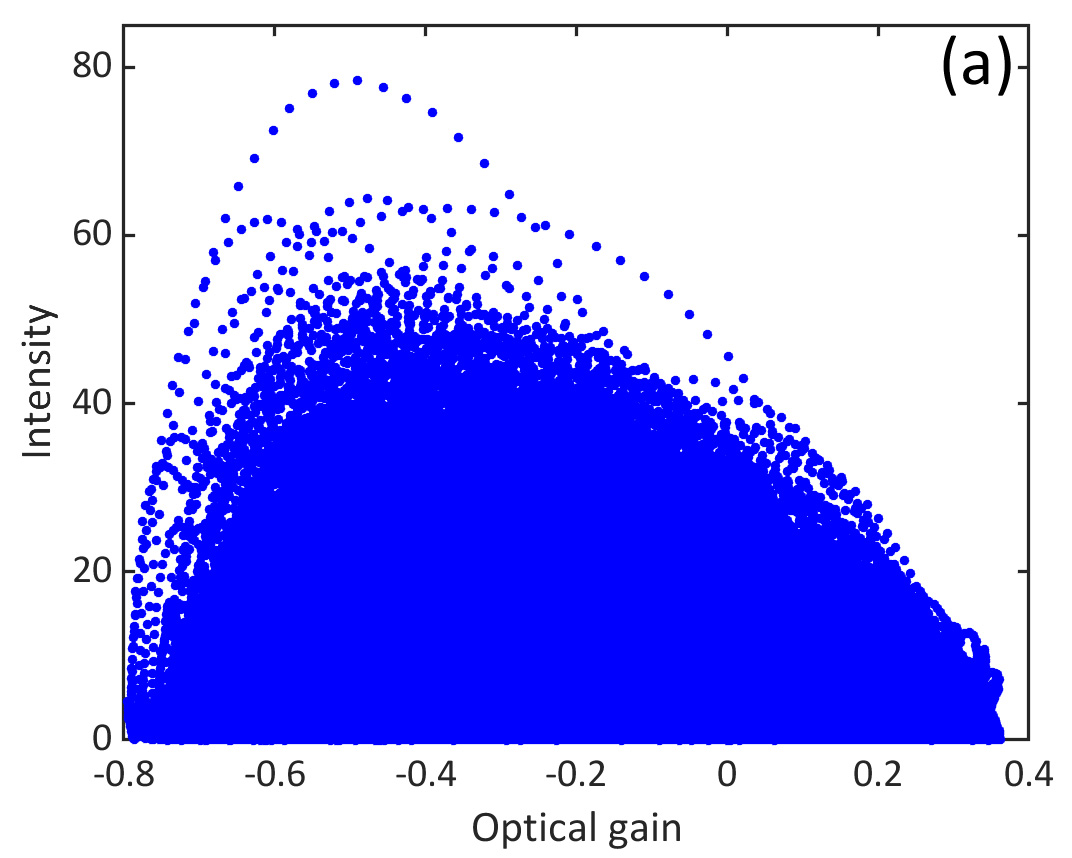} \includegraphics[width=0.49\columnwidth]{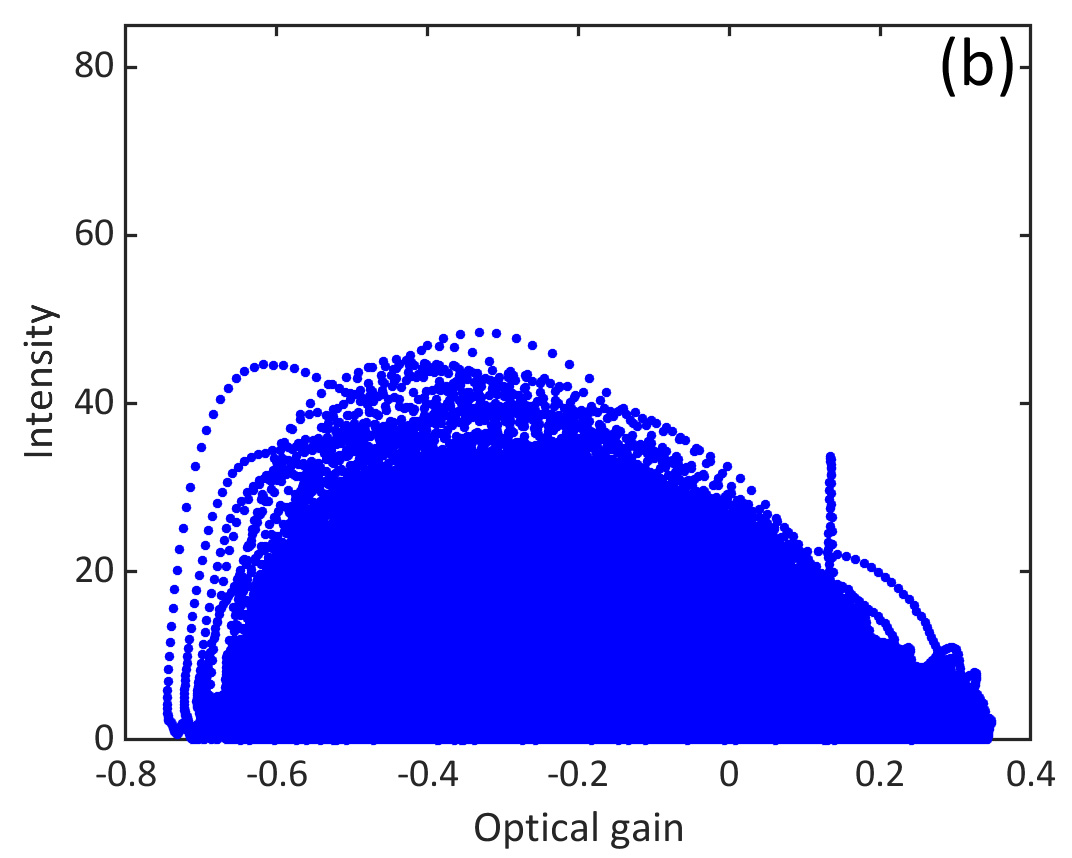}
	\caption{Trajectories of spatiotemporal maxima in sub-space made up of optical gain (D+d-1) and intensity for $\delta_1=0$ (a) and $\delta_1=0.18$ (b). Reduction of RW ratio is apparent in (b) by the smaller occupied area.  Other values are $\mu=5.0$ and $r=2.2$.}
	\label{gainInt2.2}
\end{figure}
A comparison is made between zero and a nonzero diffusion coefficient ($\delta_1=0.22$) in Fig.~\ref{RWkbeab} in terms of RW ratio and kurtosis for the cases of below ($\mu=5.0$) and above ($\mu=5.35$) laser threshold when $r$ is scanned. The figures suggest that RW ratio is essentially unaffected by carrier transverse diffusion for small $r$ values and decreases for larger $r$ values. We note that RW ratio values in Fig.~\ref{RWkbeab}(a) is almost twice those in Fig.~\ref{RWkbeab}(c) which suggests that smaller pump currents (below laser threshold) favor the formation of RWs. The same can be argued for kurtosis in Fig.~\ref{RWkbeab} (b) and (d). We should also note that turbulent state relaxes to laser off solution for larger diffusion coefficients than 0.24 at $r=2.4$ and beyond.\\
\begin{figure}
	\includegraphics[width=0.49\columnwidth]{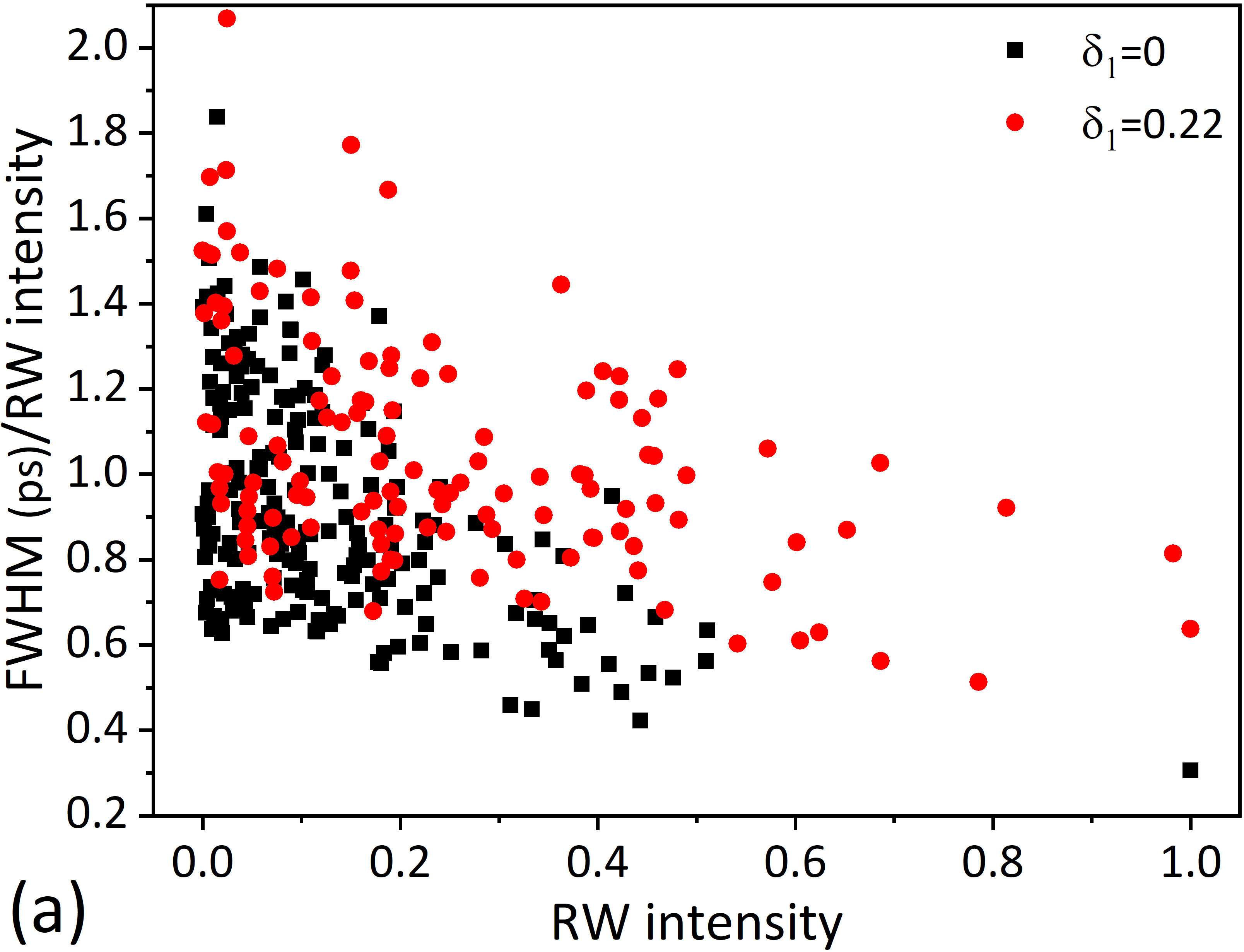} \includegraphics[width=0.49\columnwidth]{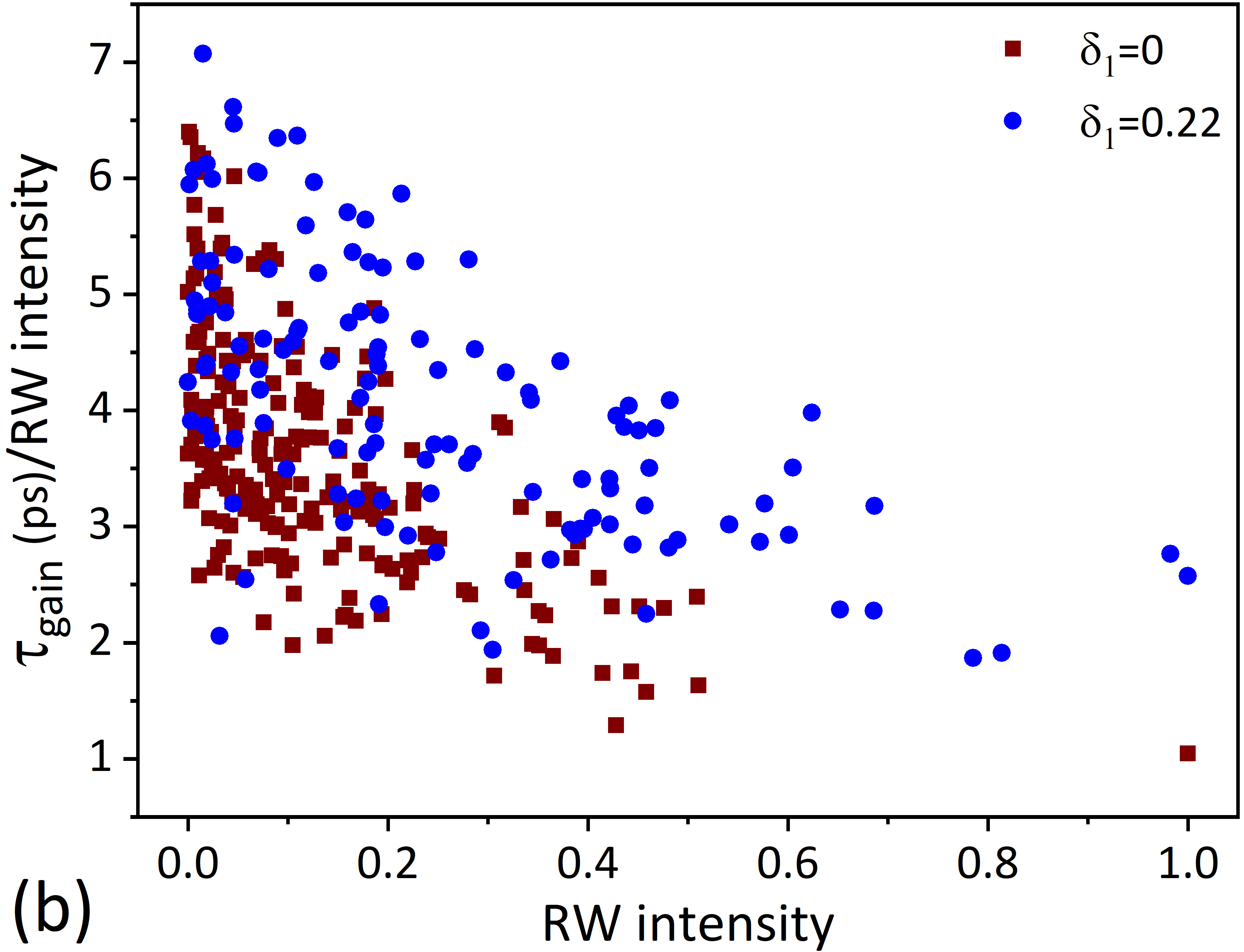}
	\caption{Temporal width of the RWs in terms of FWHM (a) and that of the corresponding optical gain $\tau_{gain}$ (b) versus normalized RW intensities compared for the cases of zero and nonzero carrier diffusion coefficient. Note that RW intensity FWHM and $\tau_{gain}$ are both normalized to the intensity of the related RW. Other values are $\mu=5.0$ and $r=2.2$.}
	\label{FWHMP}
\end{figure}
\begin{figure*}
	\includegraphics[width=0.6\columnwidth]{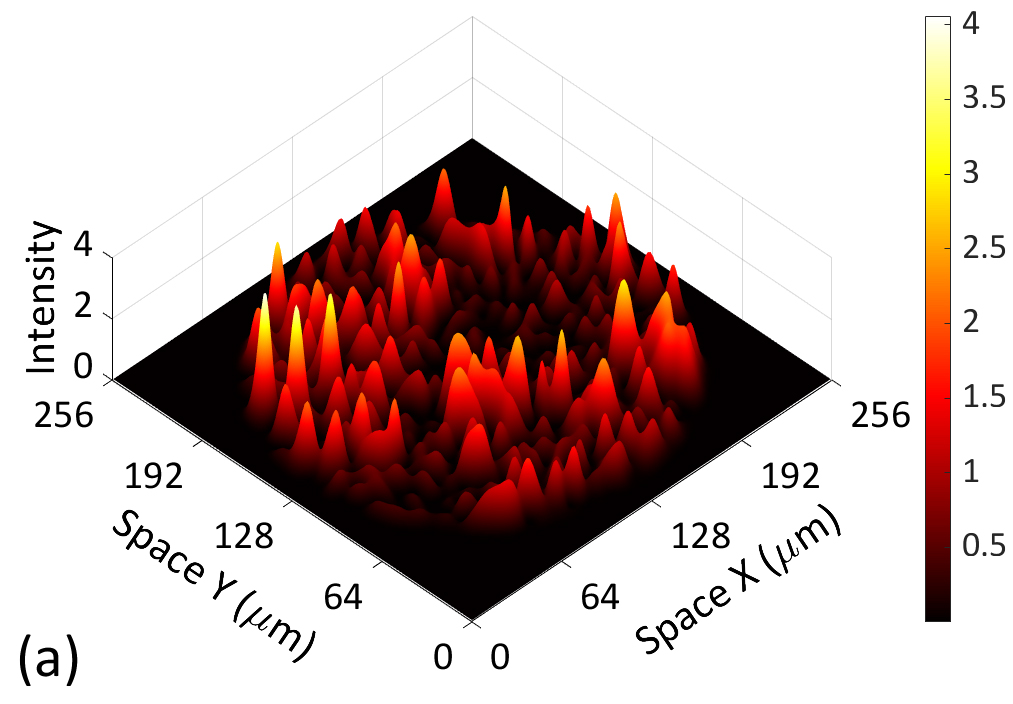}\quad \includegraphics[width=0.6\columnwidth]{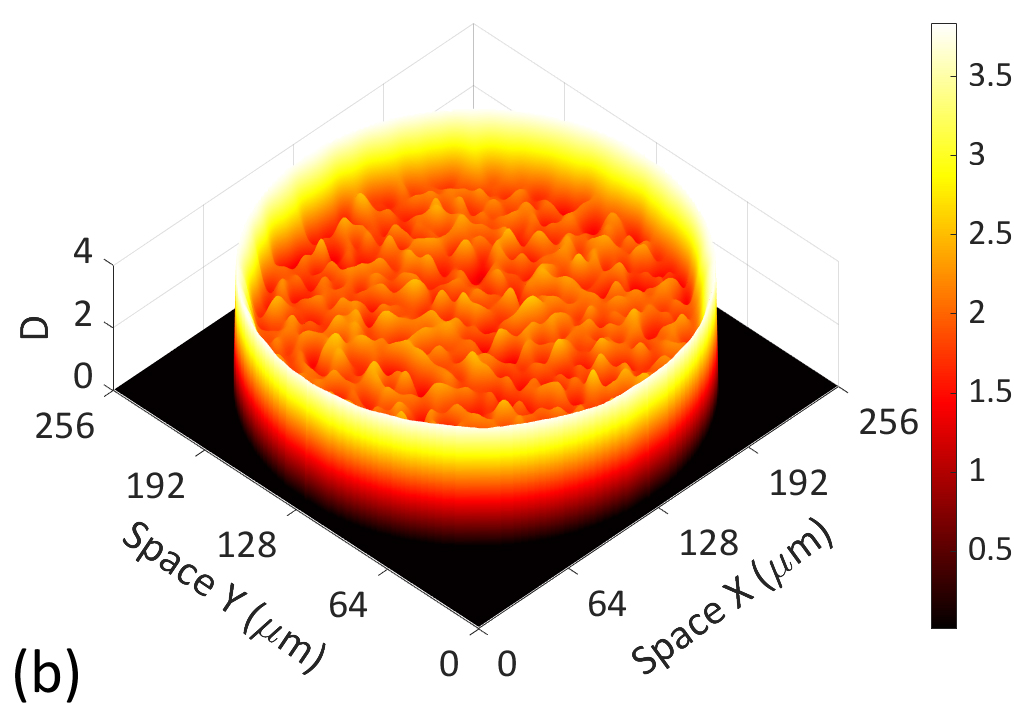}\quad \includegraphics[width=0.6\columnwidth]{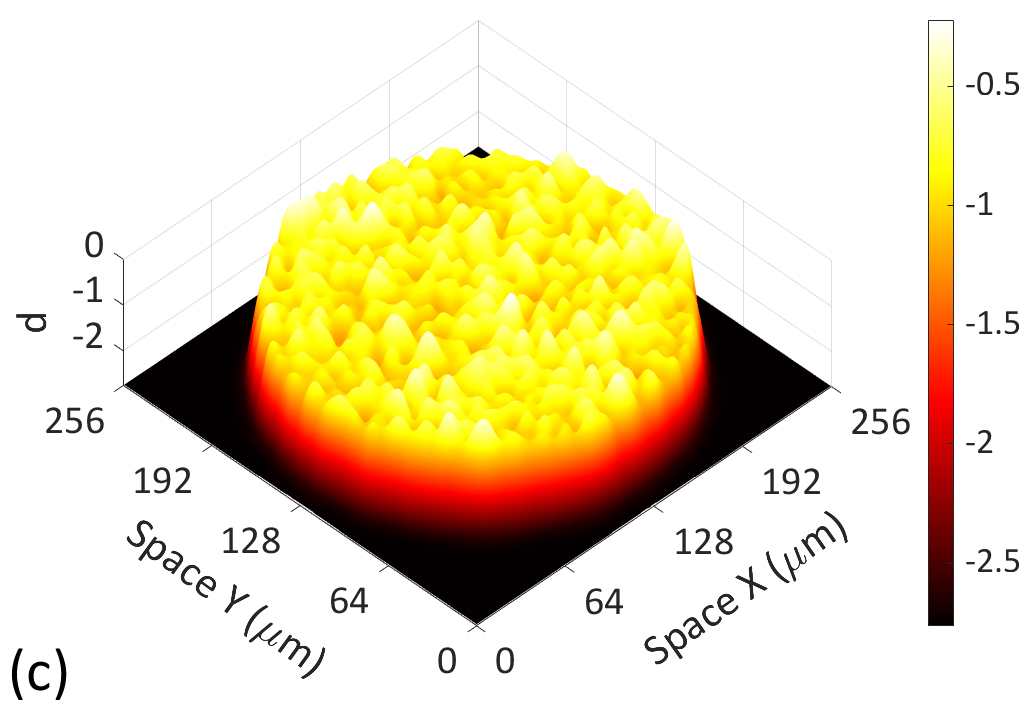}\\
	\caption{Typical transverse profiles for output intensity (a), carrier density in the amplifier (b) and that in the absorber (c) under finite current injection represented by Eq.~\ref{circeq}. $w_0$ is fixed at 120 in all the simulations discussed in this section unless stated otherwise.}
	\label{circprofs}
\end{figure*}
In Fig.~\ref{RWkdelta}, deviations of RW ratio and kurtosis from their values at zero diffusion coefficient are depicted for several values of $\delta_1$ in percentage. In these results, the pump current is kept below laser threshold at $\mu=5.0$ (left panel) and above laser threshold at $\mu=5.35$ (right panel) and several $r$ values are considered. Apart from the fact that larger transverse diffusion coefficients have more notable consequences in altering RW ratio and kurtosis values, positive and negative deviations of their values from those of zero diffusion are identified from Fig.~\ref{RWkdelta}. Positive deviations in RW ratio and kurtosis values occur for small $r$ which increases their values for nonzero diffusion coefficients, as it is evident from Fig.~\ref{RWkdelta}(a,b) and (g,h), that is, both below and above the laser threshold. The same goes for Fig.~\ref{RWkdelta}(i,j) in spite of minor irregularities. In contrast, when the value of $r$ is larger, RW ratio and kurtosis reduce and we observe negative differences from zero diffusion coefficient, as seen from figures \ref{RWkdelta}(c-f) and (k,l).\\
As an example, PDFs calculated over spatiotemporal intensity maxima along with the respective RW threshold and Weibull fits are shown in Fig.~\ref{PDF2.0} for $\mu=5.0$, $r=2.0$ and two different values of diffusion coefficients: zero and $\delta_1$ at which RW ratio increases to its maximum value according to Fig.~\ref{RWkdelta}(a,b). These specific cases are also shown in terms of their (optical gain-intensity) sub-space plots in  Fig.~\ref{gainInt2.0}. From the two figures it is evident that RW emission is enhanced in terms of both numbers and peak intensities.\\
In Fig.~\ref{RWkdelta}(c,d), we show a situation where $r$ value is larger and $\mu$ is below laser threshold meaning that carrier density is low and transverse diffusion can have a different consequence. In Fig.~\ref{PDF2.2}, we have shown PDFs calculated over spatiotemporal intensity maxima along with the respective RW thresholds and Weibull fits for $\mu=5.0$, $r=2.2$ and two different values of $\delta_1$: no diffusion and diffusion coefficient for which RW ratio reaches its minimum. The corresponding trajectories in (optical gain-intensity) sub-space plots are also shown in Fig.~\ref{gainInt2.2}. These figures confirm that nonzero diffusion coefficient reduces the number of RWs and their peak intensities when larger ratio of lifetimes is considered for carriers in active and passive media. This is in contrast with the regimes having smaller ratio of carrier lifetimes.\\
Temporal dynamics of RWs is also expected to be affected by the transverse carrier diffusion. This can be seen from the temporal width of RWs when nonzero carrier diffusion coefficient is considered. In Fig.~\ref{FWHMP}(a) normalized temporal full-width half-maximum (FWHM) values of RW intensities are depicted with respect to normalized RW intensities for $\delta_1=0$ and $\delta_1=0.22$. It is clearly seen that for nonzero carrier diffusion coefficients, RWs with the same intensity have broader temporal FWHM as a consequence of the degree of freedom corresponding to carrier drift. This is confirmed in Fig.~\ref{FWHMP}(b) where normalized temporal width of the optical gain $\tau_{gain}$ is plotted for the RWs with respect to normalized RW intensities.
\section{Finite pump effects}
\begin{figure}
	\includegraphics[width=0.49\columnwidth]{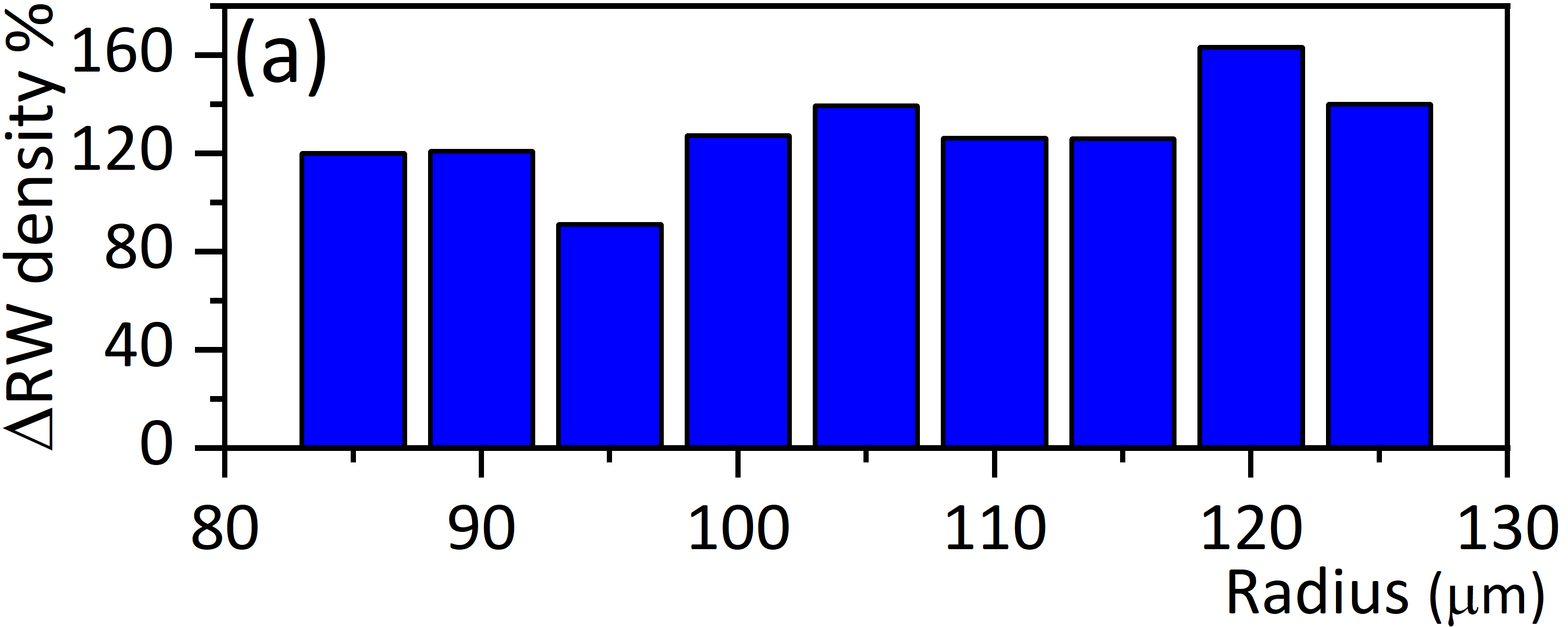} \includegraphics[width=0.49\columnwidth]{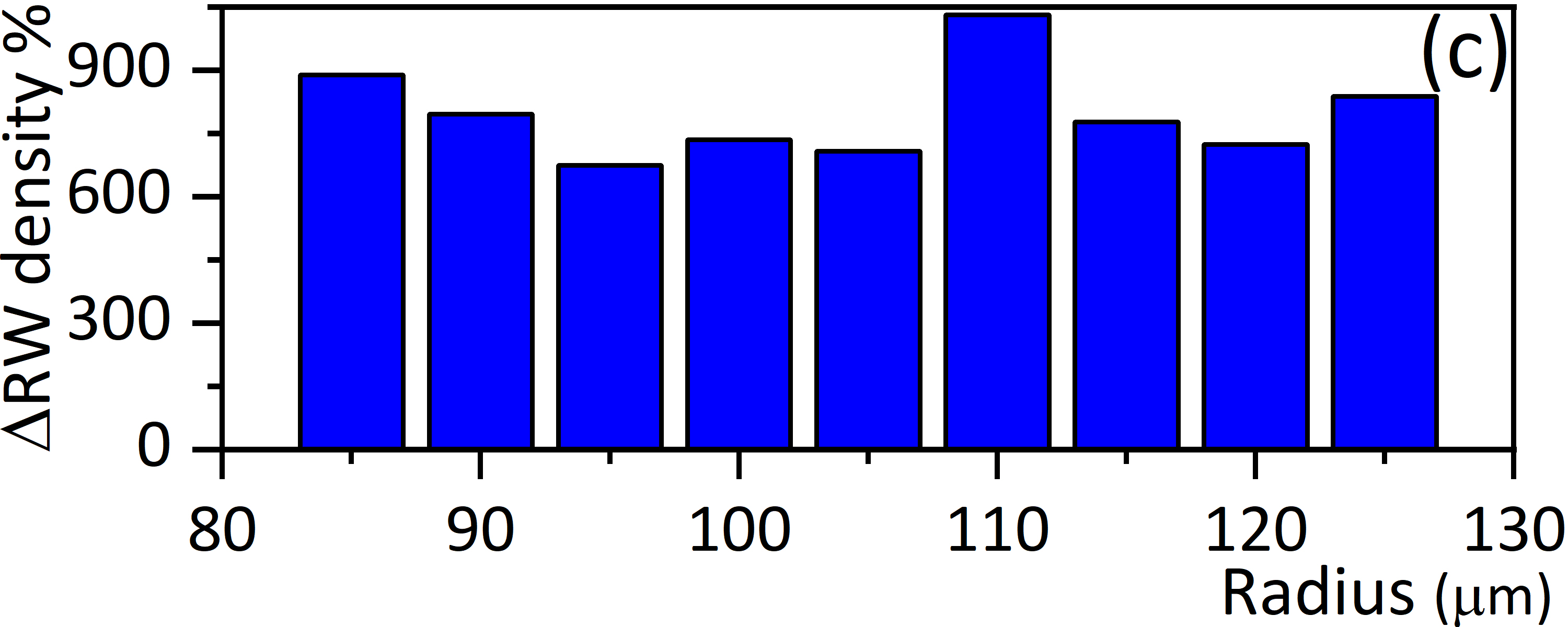}\\
	\includegraphics[width=0.49\columnwidth]{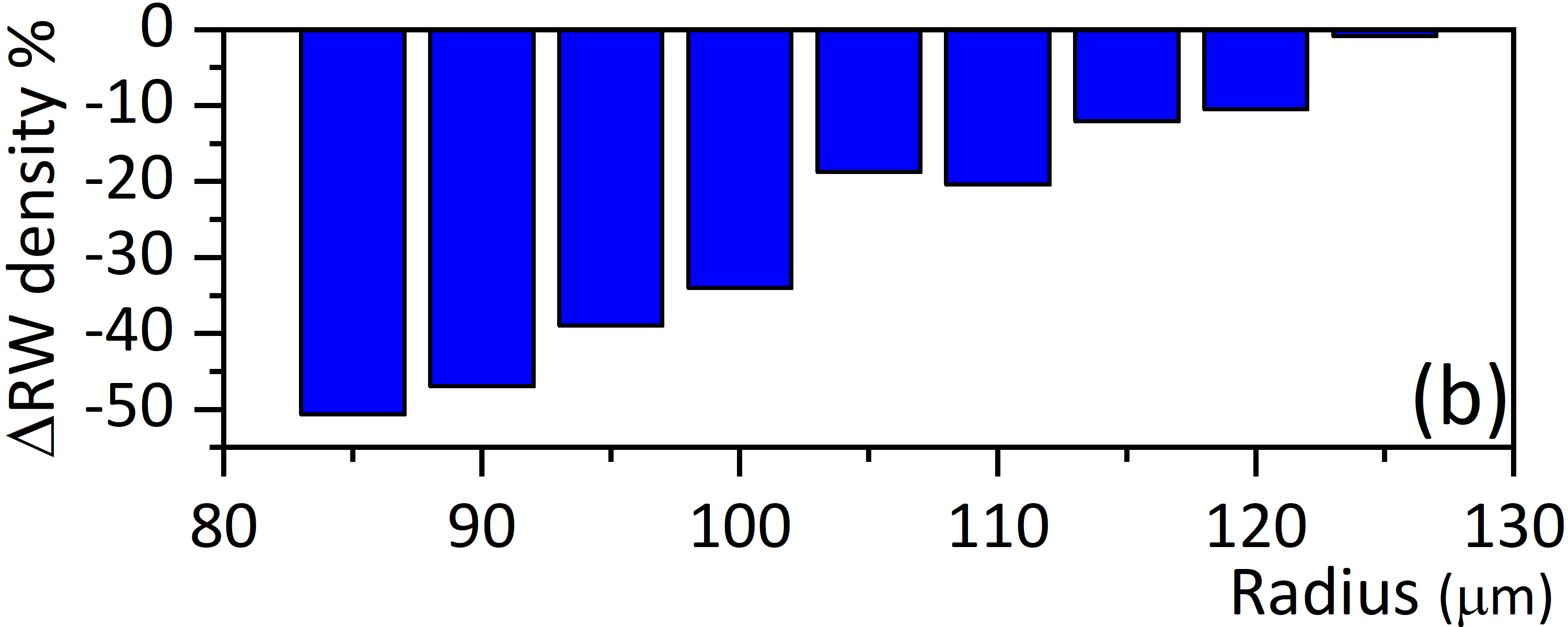} \includegraphics[width=0.49\columnwidth]{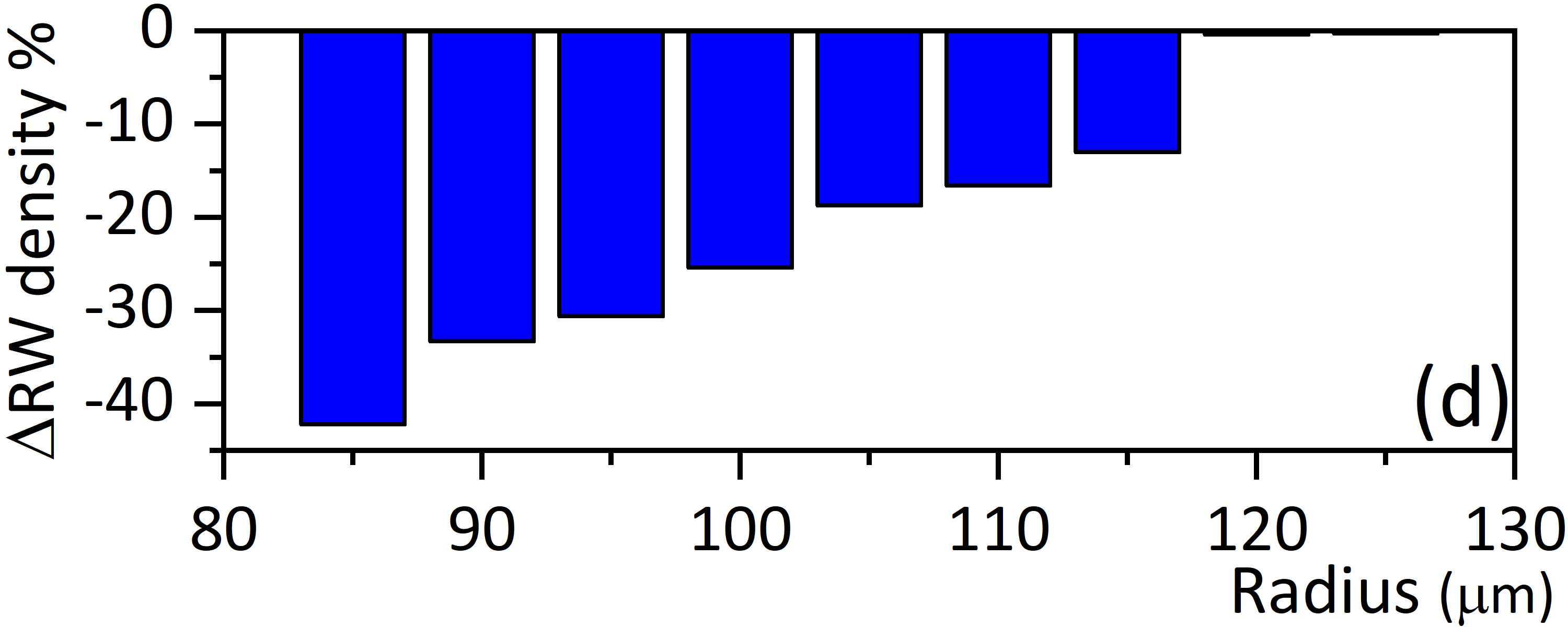}\\
	\caption{RW density difference for circular pump shape from that of infinite flat pump versus pump radius for $\delta_1=0$. The left panel shows the pump regime of below laser threshold $\mu=5.0$ and the right panel shows that of above laser threshold $\mu=5.35$. $r=2.2$ (a,c), $r=2.4$ (b) and $r=2.55$ (d).}
	\label{RWkcp}
\end{figure}

\begin{figure}
	\includegraphics[width=0.49\columnwidth]{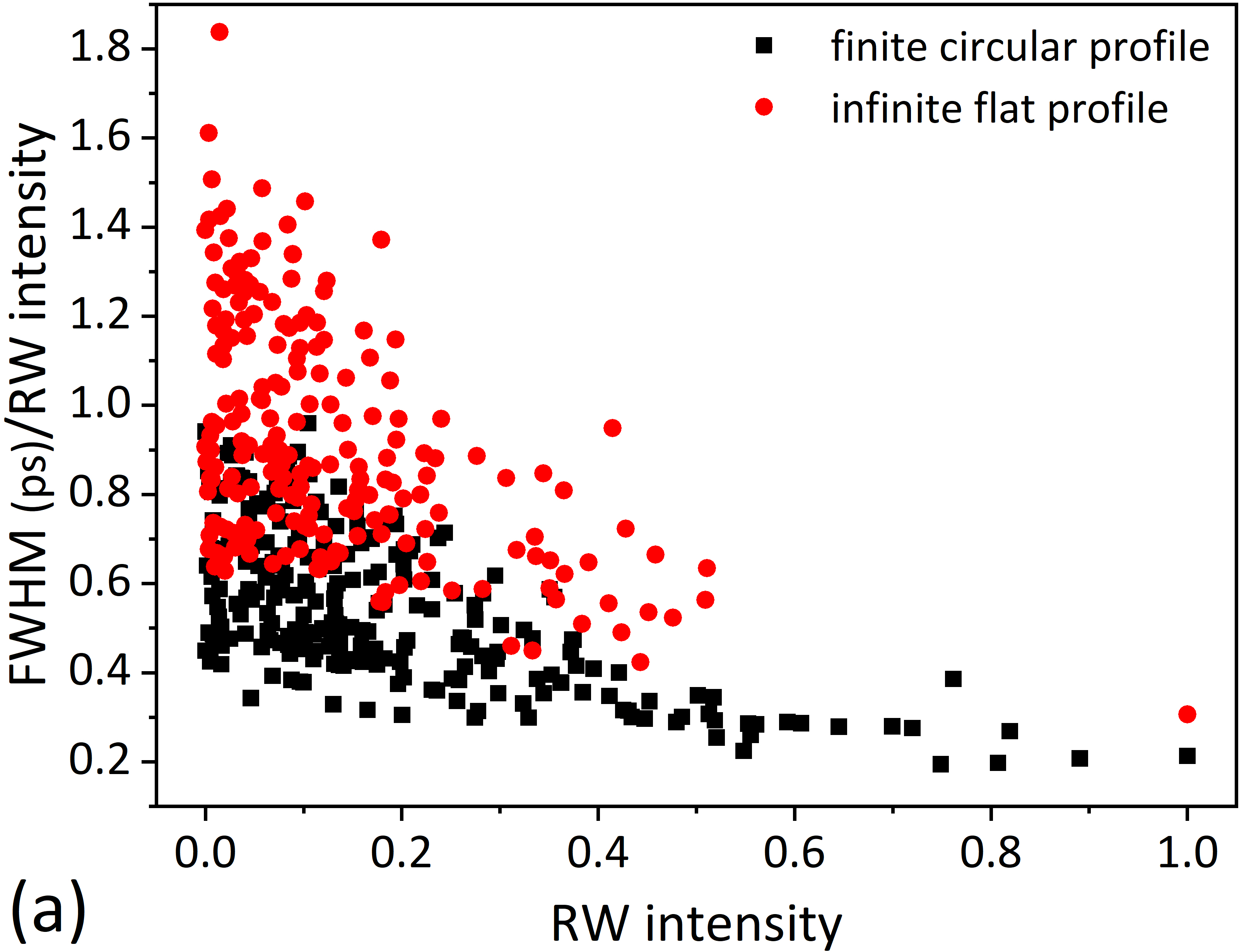} \includegraphics[width=0.49\columnwidth]{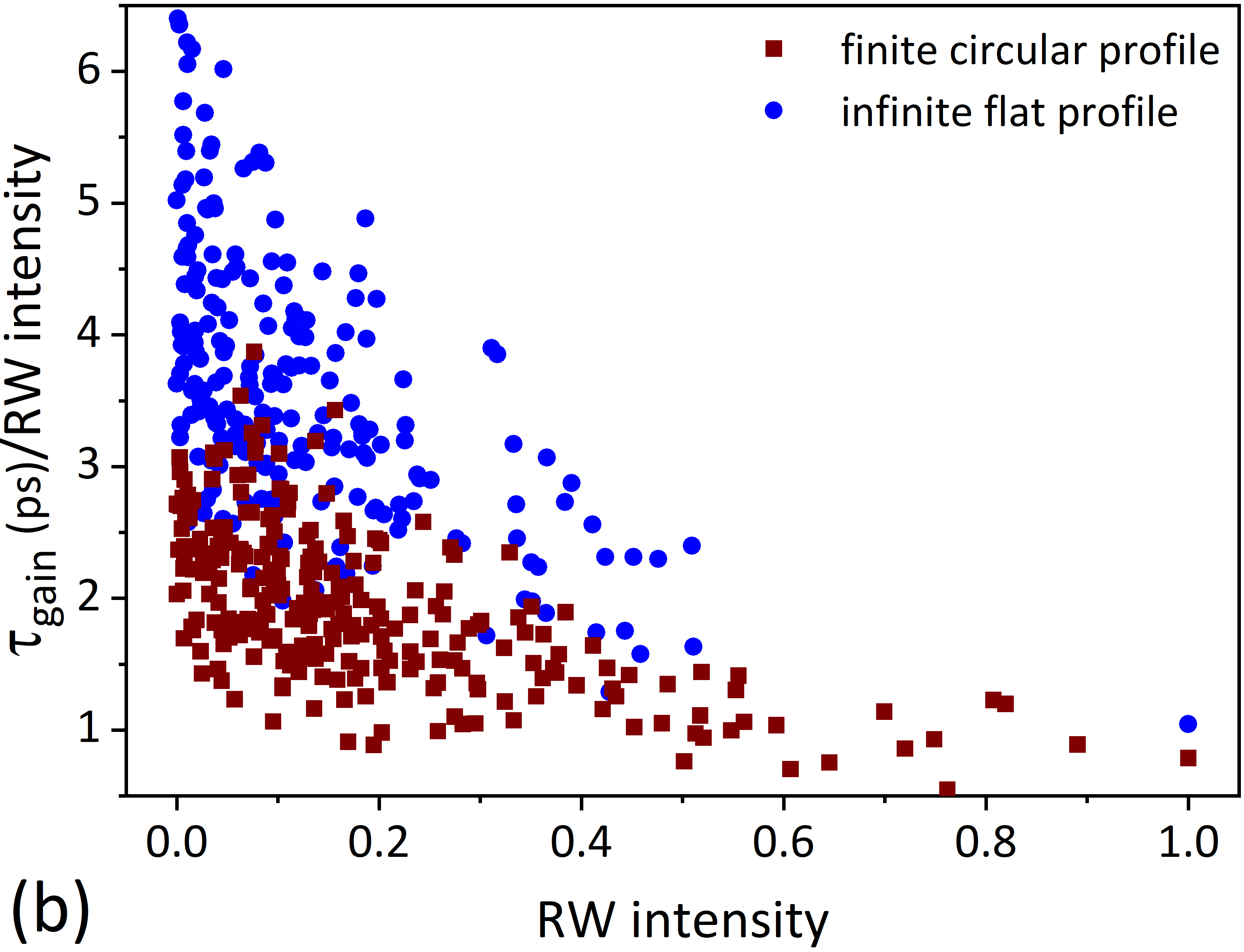}\\
	\includegraphics[width=0.49\columnwidth]{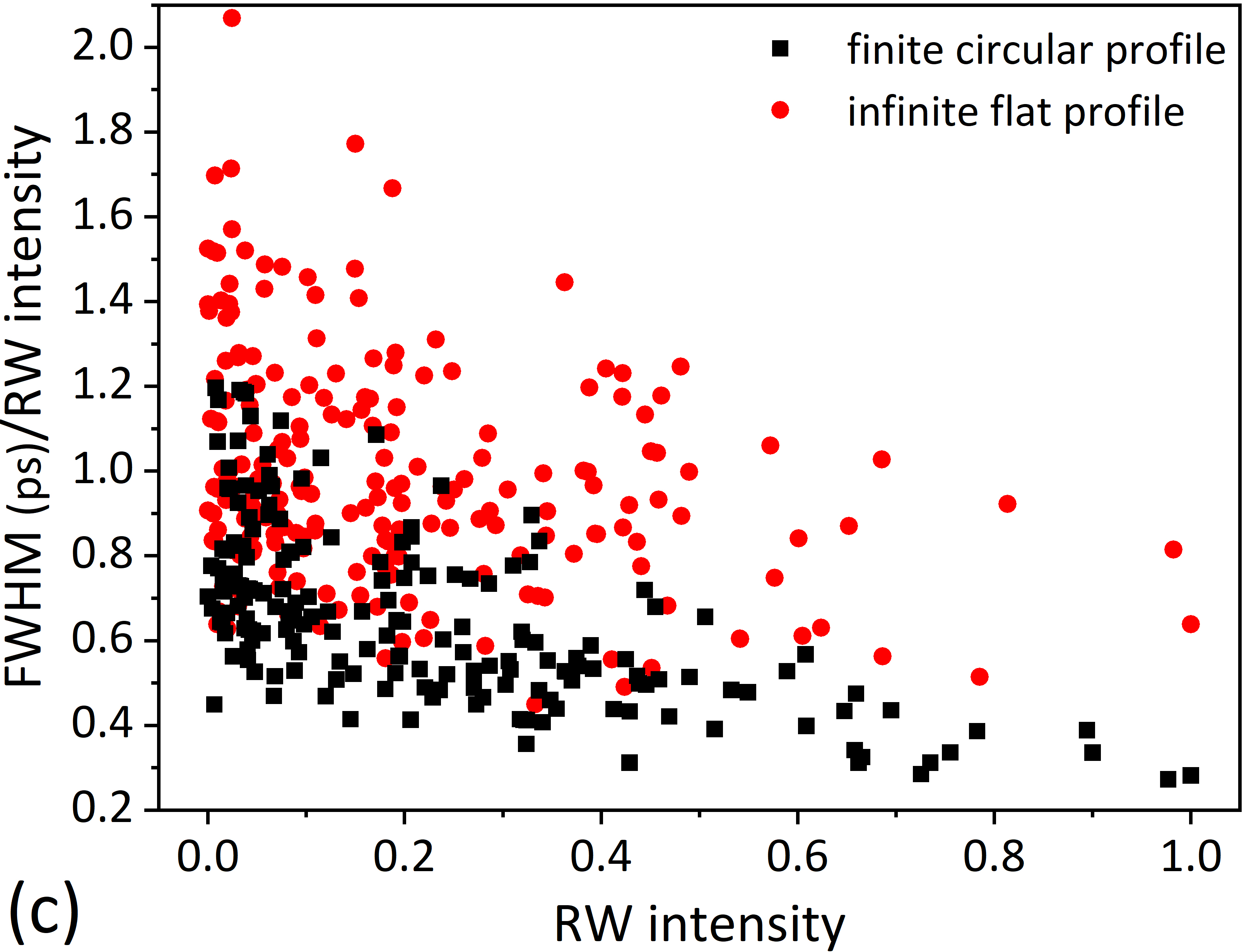} \includegraphics[width=0.49\columnwidth]{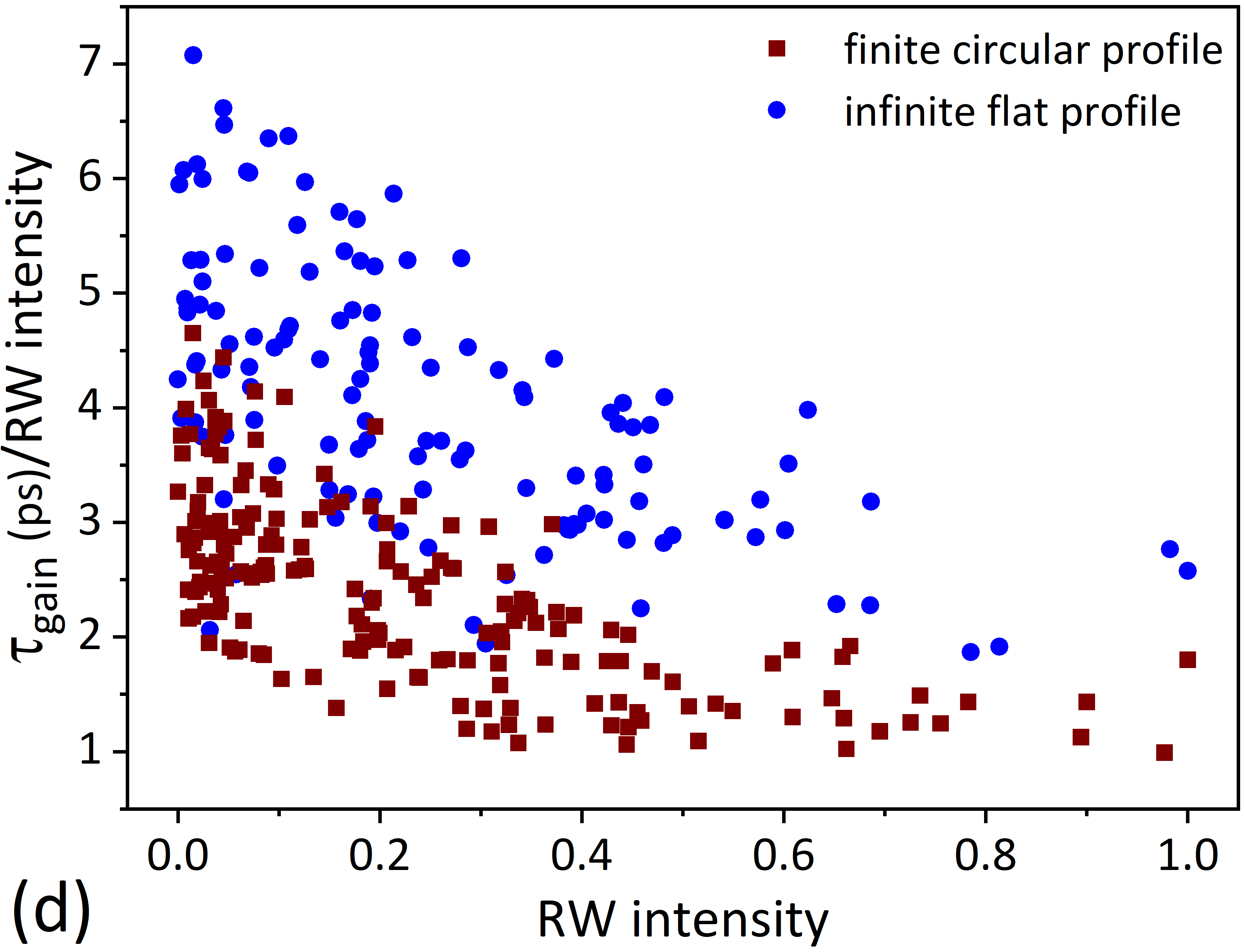}
	\caption{Temporal width of the RWs in terms of FWHM (a) and that of the corresponding optical gain $\tau_{gain}$ (b) versus normalized RW intensities compared for the cases of infinite flat pump and finite circular pump at zero carrier diffusion coefficient. (c) and (d) are the same for $\delta_1=0.22$. The ratio of the RW intensity FWHM values in finite circular pump to those in infinite flat pump on average are 0.70 and 0.58 for $\delta_1=0.22$ and $\delta_1=0$, respectively. Note that RW intensity FWHM and $\tau_{gain}$ are both normalized to the intensity of the related RW. Other values are $\mu=5.0$ and $r=2.2$.}
	\label{FWHMPC}
\end{figure}

In experimental conditions, the current-density distribution is controlled by the pump shape and is not uniform all across the transverse section. However, in numerical studies it is customary to use periodic boundary conditions which simulate an infinite pumped area. A natural and widely used pump shape in experiments is that of a circular disc which can be simulated in our model by replacing $\mu$ with
\begin{equation}
	\label{circeq}
	\mu(w)=\begin{cases}
		\mu_0, & \text{$w\leq w_0$}\\
		\mu_0 \exp{\left(-\dfrac{w-w_0}{\delta_1}\right)}, & \text{$w>w_0$}.
	\end{cases}
\end{equation}
where $\delta_1$ and $w_0$ are respectively the effective diffusion length of the injection carrier and the radius of the circular disc. $\mu_0$ is the current density within the pump area (in the top flat part) \cite{YuIEEE}. The transverse profile of the output intensity, carriers in the amplifier and absorber materials affected by the circular pump shape are shown in Fig.~\ref{circprofs}.\\
We want to check if the statistics and dynamics of RWs are altered by the inclusion of such a pump shape since the finite injection area removes the unrealistic effects of the periodic boundary condition. First, we compare infinite and finite pump shapes for $\delta_1=0$ by the deviation percentage of RW density (number of RWs per unit area) in circular pump from that in flat pump considering different circular pump radii. Fig.~\ref{RWkcp} shows that in both regimes of below and above laser threshold, RW density increases in presence of finite pump for smaller $r$ values and positive deviations are observed. However, we see negative deviations from infinite pump for larger $r$ and the values for RW density are reduced. The same comparison is made for zero and nonzero transverse carrier diffusion coefficients and is shown in Fig.~\ref{FWHMPC} for temporal dynamics of the RWs in both conditions. It is seen from Fig.~\ref{FWHMPC}(a) that infinite flat pump unrealistically gives the RWs broader temporal width at the same intensity. This fact is supported by Fig.~\ref{FWHMPC}(b) where the temporal widths of the optical gain for RWs are narrower for the same intensity when finite circular profile replaces infinite flat profile. The difference persists but reduces when a nonzero carrier diffusion coefficient is considered, see for example Fig.~\ref{FWHMPC}(c) and (d) for $\delta_1=0.22$.\\
\section{Conclusion}
We extended the results of \cite{Rimoldi17,Eslami20} by inclusion of a transverse carrier diffusion coefficient in the equations describing a broad-area semiconductor laser with a saturable absorber and performed numerical simulations to show that dynamical properties of the turbulent solutions and thus the emitted rogue waves at the peaks of spatiotemporal maxima are altered as a result of carrier transverse drift. Unlike localized structures for which the small carrier diffusion coefficient has no significant role, extended spatiotemporal structures respond to carrier transverse diffusion depending on the ratio of the carrier lifetimes in the active and passive media given by $r$. It is particularly shown that more rogue waves are emitted in presence of carrier transverse diffusion when carriers in the active material are given shorter lifetime by small $r$ value assuming that carrier lifetime in the passive material is fixed. The opposite happens for longer carrier lifetimes in the active material, larger $r$ value, and the formation of rogue waves are increasingly suppressed for larger lateral carrier diffusion coefficients. It is also illustrated that longer duration of rogue waves is followed by the nonzero transverse carrier diffusion. The importance of pump shape in experimental situations is also discussed and the statistical and dynamical behavior of rogue waves are studied under a disk-shaped pump. It is shown that less number of rogue waves per unit area for larger ratio of carrier lifetimes $r$ and more number of rogue waves per unit area for smaller $r$ values is emitted compared to the case with infinite flat pump. Moreover, it turned out that the unrealistic situation associated to infinite flat pump is responsible for longer temporal width of rogue waves since they are emitted in shorter duration under a finite pump.

\end{document}